\documentclass{article}

\PassOptionsToPackage{backref=page}{hyperref}

\usepackage{microtype}
\usepackage{graphicx}
\usepackage{subcaption}
\usepackage{booktabs} 

\usepackage{xspace}
\usepackage{upgreek}

\usepackage{hyperref}

\usepackage[accepted]{icml2024} 

\usepackage{amsmath}
\usepackage{amssymb}
\usepackage{mathtools}
\usepackage{amsthm}

\usepackage[capitalize,noabbrev]{cleveref}

\theoremstyle{plain}

\theoremstyle{definition}

\theoremstyle{remark}

\usepackage[textsize=tiny]{todonotes}

\makeatletter
\DeclareRobustCommand\onedot{\futurelet\@let@token\@onedot}
\def\@onedot{\ifx\@let@token.\else.\null\fi\xspace}

\def\eg{\emph{e.g}\onedot, } 
\def\ie{\emph{i.e}\onedot, }

\makeatother

\usepackage{amsmath,amsfonts,bm}

\def\eqref#1{equation~\ref{#1}}

\def\1{\bm{1}}

\def\rlambda{{\textnormal{$\uplambda$}}}

\def\rvepsilon{{\bm{\upepsilon}}}

\def\rvmu{{\bm{\upmu}}}

\def\rvf{{\mathbf{f}}}

\def\rvn{{\mathbf{n}}}

\def\rvv{{\mathbf{v}}}

\def\rvx{{\mathbf{x}}}
\def\rvy{{\mathbf{y}}}
\def\rvz{{\mathbf{z}}}

\def\vy{{\bm{y}}}

\def\mD{{\bm{D}}}

\def\mI{{\bm{I}}}

\def\mP{{\bm{P}}}

\DeclareMathAlphabet{\mathsfit}{\encodingdefault}{\sfdefault}{m}{sl}
\SetMathAlphabet{\mathsfit}{bold}{\encodingdefault}{\sfdefault}{bx}{n}

\def\gN{{\mathcal{N}}}

\newcommand{\E}{\mathbb{E}}

\newcommand{\Cov}{\mathrm{Cov}}

\newcommand{\urlofwebpage}[0]{https://hilamanor.github.io/AudioEditing/}
\newcommand{\urlofwebpagesupp}[0]{https://hilamanor.github.io/AudioEditing/supp.html}

\icmltitlerunning{Zero-Shot Unsupervised and Text-Based Audio Editing Using DDPM Inversion}

\begin{document}

\twocolumn[
\icmltitle{Zero-Shot Unsupervised and Text-Based Audio Editing Using DDPM Inversion}



\icmlsetsymbol{equal}{*}

\begin{icmlauthorlist}
\icmlauthor{Hila Manor}{tech}
\icmlauthor{Tomer Michaeli}{tech}
\end{icmlauthorlist}

\icmlaffiliation{tech}{Faculty of Electrical and Computer Engineering, Technion -- Israel Institute of Technology, Haifa, Israel}
\icmlcorrespondingauthor{Hila Manor}{hila.manor@campus.technion.ac.il}

\icmlkeywords{Machine Learning, 
Editing, Diffusion models, Audio editing, Music Editing, Supervised editing, Unsupervised editing, Prompt-based editing, Text-guided editing, Text-based editing, DDPM inversion}

\vskip 0.3in
]



\printAffiliationsAndNotice{}  

\begin{abstract}

Editing signals using large pre-trained models, in a zero-shot manner, has recently seen rapid advancements in the image domain. 
However, this wave has yet to reach the audio domain. 
In this paper, we explore two zero-shot editing techniques for audio signals, which use DDPM inversion with pre-trained diffusion models.
The first, which we coin \emph{ZEro-shot Text-based Audio (ZETA)} editing, is adopted from the image domain. The second, named \emph{ZEro-shot UnSupervized (ZEUS)} editing, is a novel approach for discovering semantically meaningful editing directions without supervision. 
When applied to music signals, this method exposes a range of musically interesting modifications, from controlling the participation of specific instruments to improvisations on the melody.
Samples and code can be found on our \href{\urlofwebpage}{examples page}.
\end{abstract}

\section{Introduction}

\begin{figure*}[t]
    \centering
    \includegraphics[width=\linewidth]{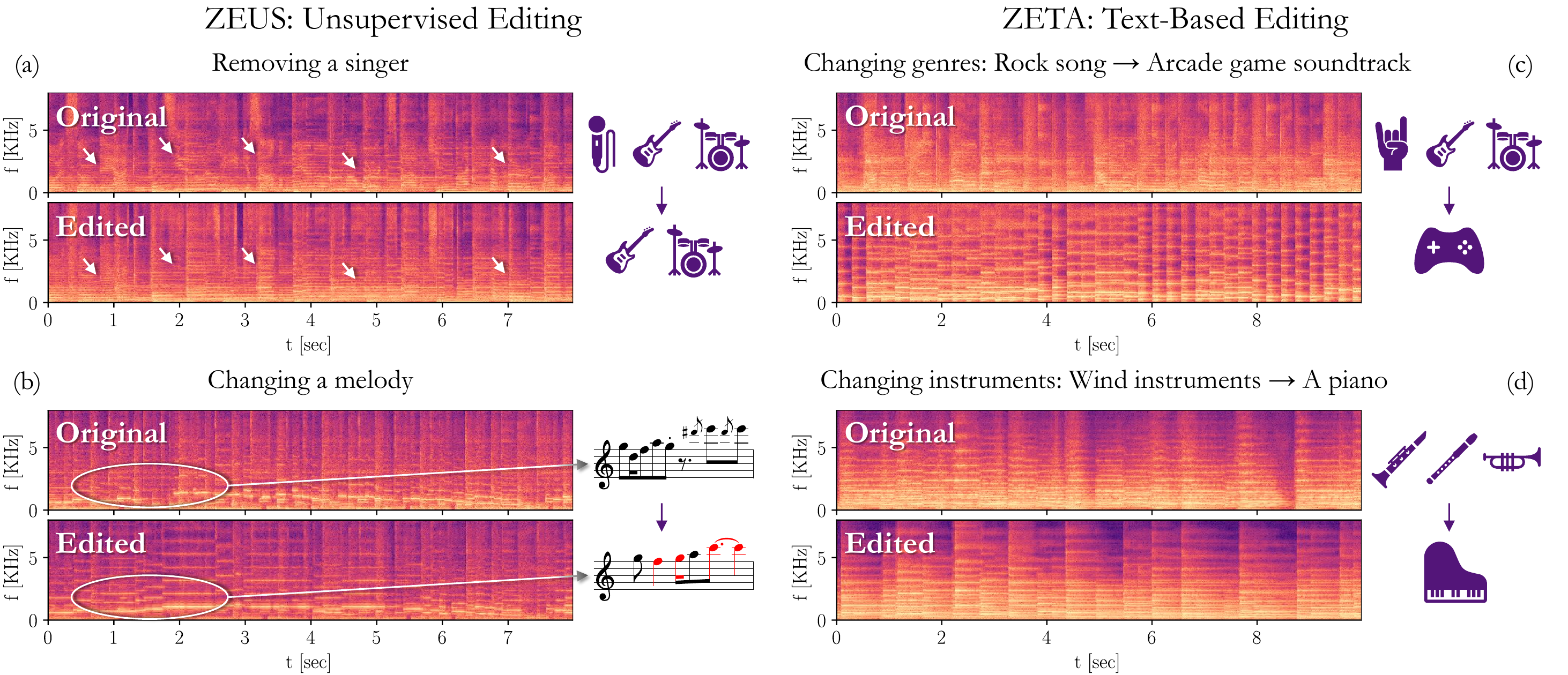}
    \caption{
    \textbf{Zero-shot audio editing.} We present two methods for editing audio signals using DDMs: ZEUS, a novel unsupervised approach (left), and ZETA, a text-based approach adopted from the image domain (right). Both methods can edit a variety of concepts from style to instrumentation. (a) The singer (curved pitches) is removed while the rest of the signal remains intact. (b) The melody notes change, reflected by a change in the dominant pitch. (c) The genre is changed, affecting the entire statistics of the spectrogram. (d) The instrumentation changes from a woodwind section to a piano, visible by the attack (abrupt starts) of the piano keys. All examples can be listened to in our \href{\urlofwebpage}{examples page}.
    For (c),(d), $T_\text{start}=100,70$, respectively (Sec.~\ref{sec:supervisedEditing}). For (a),(b), $T_\text{start}=150, 200$, $t'=115,80$, $T_\text{end}=1$, using the top 3 PCs (Sec.~\ref{sec:unsupervisedEditing}).}
    \label{fig:teaser}
\end{figure*}

\begin{figure*}[t]
    \centering
    \includegraphics[width=\linewidth]{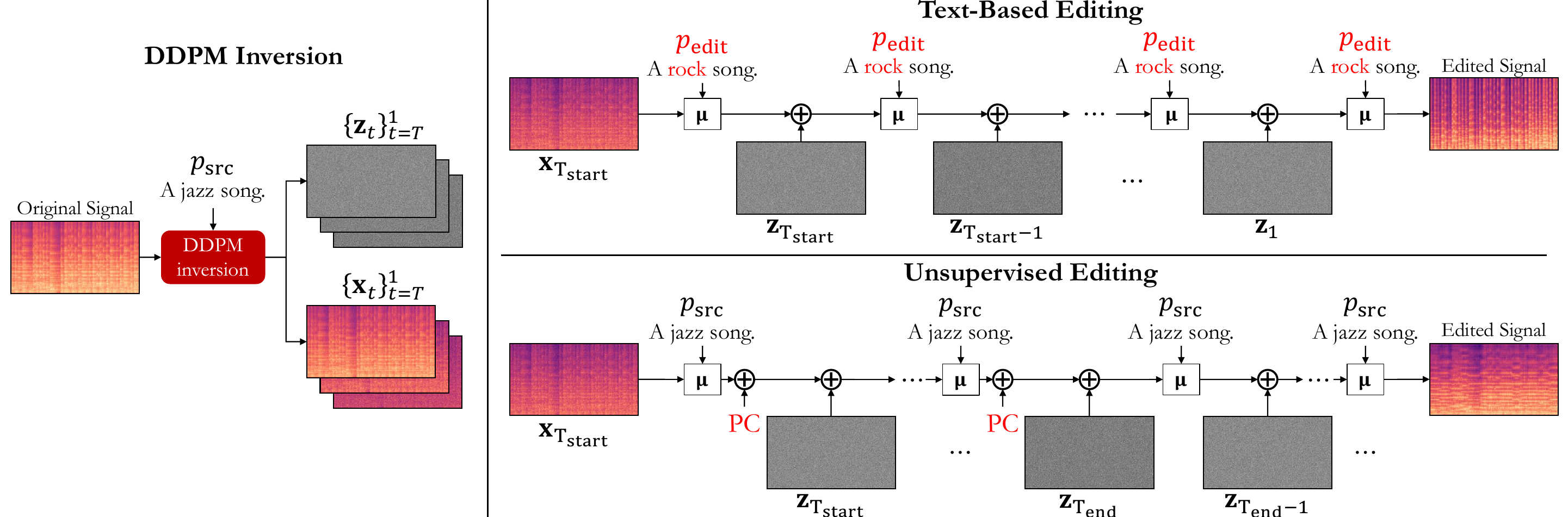}
    \caption{\textbf{Overview of our text-based and unsupervised editing methods.} We start by extracting the noise vectors corresponding to an input signal using DDPM inversion, optionally conditioned on a text prompt $p_\text{src}$. For the text-based editing approach, we then continue the reverse process with a different text prompt. For the unsupervised approach, we continue the reverse process when applying PCs calculated on the forward process. Red color shows what changed in the reverse process.} 
    \label{fig:scheme}
\end{figure*}

Creative media creation has seen a dramatic transformation with the recent advancements in text-based generative models, particularly those based on denoising diffusion models (DDMs) \citep{sohl2015deep, ho2020denoising, song2020denoising}. While progress has been initially made in image synthesis \cite{ramesh2021zero,rombach2022high}, generative models for the audio domain have recently captured increased interest. Indeed, transformer based models such as MusicGen~\citep{copet2023simple} and MusicLM~\citep{agostinelli2023musiclm}, and diffusion based models such as AudioLDM~\citep{liu2023audioldm} and TANGO~\citep{ghosal2023text}, now enable nonprofessional users to effortlessly create short musical excerpts and audio effects.

To allow more fine-grained manipulations, a lot of attention has been recently devoted to \emph{editing} of signals using DDMs. In the image domain, some works proposed to train from scratch text-guided models for editing~\citep{brooks2023instructpix2pix}, or to use test-time optimization to control the generation or to fine-tune a pre-trained text-to-image model~\citep{gal2022image, kim2022diffusionclip, kawar2023imagic, ruiz2023dreambooth, zhang2023sine}. 
Other works demonstrated that high quality results can be obtained with zero-shot editing methods that employ pre-trained text-to-image models~\citep{meng2021sdedit, huberman2023edit, tumanyan2023plug, wu2023latent}, avoiding the heavy computational burden of test-time optimization. In the audio domain, text-based editing has only very recently started gaining traction. Recent works either train models from scratch for specific editing tasks~\citep{copet2023simple, han2023instructme, wang2023audit}, or apply test-time optimization~\citep{paissan2023audio, plitsis2023investigating}.
To date, zero-shot editing for audio signals has only been illustrated in the AudioLDM work~\citep{liu2023audioldm}, using the naive SDEdit~\citep{meng2021sdedit} approach.

In this paper we explore two approaches for zero-shot audio editing with pre-trained audio DDMs, one based on \emph{text guidance} and the other based on semantic perturbations that are found in an \emph{unsupervised} manner. 
Our zero-shot text-guided audio (ZETA) editing technique allows a wide range of manipulations, from changing the style or genre of a musical piece to changing specific instruments in the arrangement (Fig.~\ref{fig:teaser}(c),(d)), all while maintaining high perceptual quality and semantic similarity to the source signal. 
Our zero-shot unsupervised (ZEUS) technique allows generating \eg interesting variations in melody that adhere to the original key, rhythm, and style, but are impossible to achieve through text guidance (Fig.~\ref{fig:teaser}(a),(b)).

Our methods are based on the recently introduced edit-friendly DDPM inversion method~\citep{huberman2023edit}, which we use for extracting latent noise vectors corresponding to the source signal. To generate the edited signal, we use those noise vectors in a DDPM sampling process~\citep{ho2020denoising}, while drifting the diffusion towards the desired edit. 
In our text-based method, we achieve this by changing the text prompt supplied to the denoiser model. In our unsupervised method, we perturb the output of the denoiser in the directions of the top principal components (PCs) of the posterior, which we efficiently compute based on \citet{manor2023posterior}. As we show, these perturbations are particularly useful for editing music excerpts, in which they can uncover improvisations and other musically plausible modifications.

We compare our methods to the state-of-the-art text-to-music model MusicGen~\citep{copet2023simple}, whose generation process can be conditioned on a given music piece, as well as to using the zero-shot editing methods SDEdit~\citep{meng2021sdedit} and DDIM inversion~\citep{song2021denoising, dhariwal2021diffusion} in conjunction with the AudioLDM2 model~\citep{liu2023audioldm2}. 
We show that our approaches outperforms these methods in terms of generating semantically meaningful modifications, while remaining faithful to the original signal's structure.

\section{Related Work}\label{sec:relatedwork}

\paragraph{Specialized audio editing models.}
The most common approach for editing audio is to train specialized models for this particular task.
MusicGen~\citep{copet2023simple} and MusicLM~\citep{agostinelli2023musiclm} train language based models for generating audio conditioned on text, and optionally also on a given melody.
Editing a music excerpt with MusicGen is achieved by conditioning the generation on the excerpt's chromagram while supplying a text prompt describing the desired edit.
However, because of its reliance on chromagrams, it typically fails in editing polyphonic music.
MusicLM conditioning is built on a novel proprietary joint music-text embedding space, named MuLan, built to encode monophonic melodies.
A different approach, borrowed from the image domain \cite{brooks2023instructpix2pix}, is to train an instruction based diffusion model for editing. This has been done for general audio~\citep{wang2023audit} as well as specifically for music~\citep{han2023instructme}. 
These methods are limited to a small set of modifications (\eg ``Add'', ``Remove'', ``Replace'') and require training on a large dataset of triplets (text prompt, input audio, and output audio).
Our methods require no training and are not limited to a fixed set of instructions.

\paragraph{Test-time optimization.}
Instead of training a model from scratch, some works leverage large pre-trained models for editing. \citet{paissan2023audio} and \citet{plitsis2023investigating} demonstrated the effectiveness of test-time optimization methods, adopted from the image domain~\citep{gal2022image, kawar2023imagic, ruiz2023dreambooth}, to editing of audio signals.
These methods either fine-tune the diffusion model to reconstruct the given signal~\citep{ruiz2023dreambooth}, optimize the text-embedding to reconstruct the signal~\citep{gal2022image}, fine-tune the latent noise vector using some feature matching loss~\cite{novack2024ditto}, or use some combination of these approaches~\citep{kawar2023imagic}. However, performing optimization for each new signal at test-time is computationally intensive. 
Moreover, these methods struggle with changing specific concepts, \eg replacing only the piano in a music piece with a banjo. Our techniques avoid test-time optimization, and can achieve focused edits.

\paragraph{Zero-shot editing.}
Some works focus on zero-shot editing using pre-trained diffusion models. Perhaps the simplest approach is SDEdit~\citep{meng2021sdedit}, which adds noise to the signal and then runs it through the reverse diffusion process with a different text prompt. SDEdit was recently used for audio~\citep{liu2023audioldm} as well as for piano-roll music~\citep{zhang2023sdmuse}. However it suffers from a severe tradeoff between adherence to the text and adherence to the original signal. Another direction, which has become popular in the image domain, is to use inversion techniques that extract the diffusion noise vectors corresponding to the source signal. One method for doing so is DDIM inversion~\citep{song2021denoising, dhariwal2021diffusion}. This method was found suboptimal for editing images on its own, and is therefore typically accompanied by intervention in the attention maps during the diffusion process~\citep{hertz2022prompt, cao2023masactrl, tumanyan2023plug}. 
A concurrent work by \citet{zhang2024musicmagus} recently proposed to edit audio by using DDIM inversion combined with latent space manipulations. However, inline with our observations about DDIM inversion, their method's applicability to real audio signals is limited.
Another approach is DDPM inversion \citep{huberman2023edit,wu2023latent}, which is conceptually similar, but applies to the DDPM sampling scheme. 
Here we adopt the DDPM inversion method of~\citet{huberman2023edit}, which has shown state-of-the-art results in the image domain.

\paragraph{Unsupervised editing.}
Finding semantic editing directions in an unsupervised manner, without any guidance or reference samples, has been exhaustively studied in the context of GANs~\citep{spingarn2020gan, shen2020interpreting, Shen_2021_CVPR, wu2021stylespace}. Recently, several works proposed ways for finding editing directions in the bottleneck features ($h$-space)~\citep{kwon2022diffusion} of a diffusion model~\citep{haas2023discovering, park2023unsupervised, jeong2024training} in an unsupervised manner. 
The unsupervised method we explore in this paper finds editing directions in the noise space of the diffusion model. This is done through adaptation of the method of \citet{manor2023posterior}, which quantifies uncertainty in Gaussian denoising.

\section{Method}

\subsection{DDPM Inversion}\label{sec:DDPM_def}

Denoising diffusion probabilistic models (DDPMs) \citep{ho2020denoising} generate samples through an iterative process, which starts with a Gaussian noise vector $\rvx_T\sim\gN(0,\mI)$ and gradually denoises it in $T$ steps as
\begin{equation}\label{eq:reverseProcess}
    \rvx_{t-1}= \rvmu_t(\rvx_t) + \sigma_t \rvz_t,\quad t=T,\ldots,1.
\end{equation}
Here, $\{\rvz_t\}$ are iid standard Gaussian vectors, $\{\sigma_t\}$ is an increasing sequence of noise levels, and $\rvmu_t(\rvx_t)$ is a linear function of $\hat{\rvx}_{0|t}$, which is the MSE-optimal prediction of a clean signal $\rvx_0$ from its noisy version 
\begin{equation}\label{eq:forwardProcess}
    \rvx_t = \sqrt{\bar{\alpha}_t}\rvx_0+\sqrt{1-\bar{\alpha}_t}\rvepsilon_t, \quad \rvepsilon_t\sim\gN(0,\mI).
\end{equation}
The coefficients $\{\bar{\alpha}_t\}$ monotonically decrease from $1$ to $0$.

Rather than generating a synthetic signal, here we are interested in editing a real audio excerpt, $\rvx_0$. To do so, we follow the general approach of \citet{huberman2023edit} and \citet{wu2023latent}. Specifically, we start by extracting noise vectors  $\{\rvx_T, \rvz_T,...,\rvz_1\}$ that cause the sampling process (\ref{eq:reverseProcess}) to generate the given signal $\rvx_0$ at $t=0$. This is called \emph{inversion}. We then use those noise vectors to sample a signal using (\ref{eq:reverseProcess}) while steering the generation towards a desired edit effect, as we detail in Sections~\ref{sec:supervisedEditing} and~\ref{sec:unsupervisedEditing}. 
To extract the noise vectors, we use the edit-friendly DDPM inversion method of \citet{huberman2023edit}. This method accepts as input the source signal $\rvx_0$ and generates from it an auxiliary sequence of vectors
\begin{equation}\label{eq:xt_inversion}
    \rvx_t=\sqrt{\bar{\alpha}_t}\rvx_0 +\sqrt{1-\bar{\alpha}_t}\tilde{\rvepsilon}_t, \quad t=1,\ldots,T, 
\end{equation}
where $\tilde{\rvepsilon}_t$ are sampled independently from $\gN(0,\mI)$. It then extracts the noise vectors by isolating them from (\ref{eq:reverseProcess}) as
\begin{equation}\label{eq:zt_inversion}
    \rvz_t=(\rvx_{t-1}-\rvmu_t(\rvx_t))/\sigma_t, \quad t=T,\ldots,1.
\end{equation}
While the noise vectors constructed this way have a different distribution than those participating in the original generative process (\ref{eq:reverseProcess}), they have been shown to encode the global structure of the signal $\rvx_0$ more strongly, making them particularly suitable for editing tasks.

We note that a diffusion process can be either performed in the raw waveform space or in some latent space \citep{rombach2022high}. In this work we utilize the pre-trained AudioLDM2~\citep{liu2023audioldm, liu2023audioldm2} model, which works in a latent space. AudioLDM2 generates mel-spectograms conditioned on text. Those mel-spectograms are decoded into waveforms using HiFi-GAN~\citep{kong2020hifi}.

\begin{figure*}[t]
    \centering
    \includegraphics[width=\linewidth]{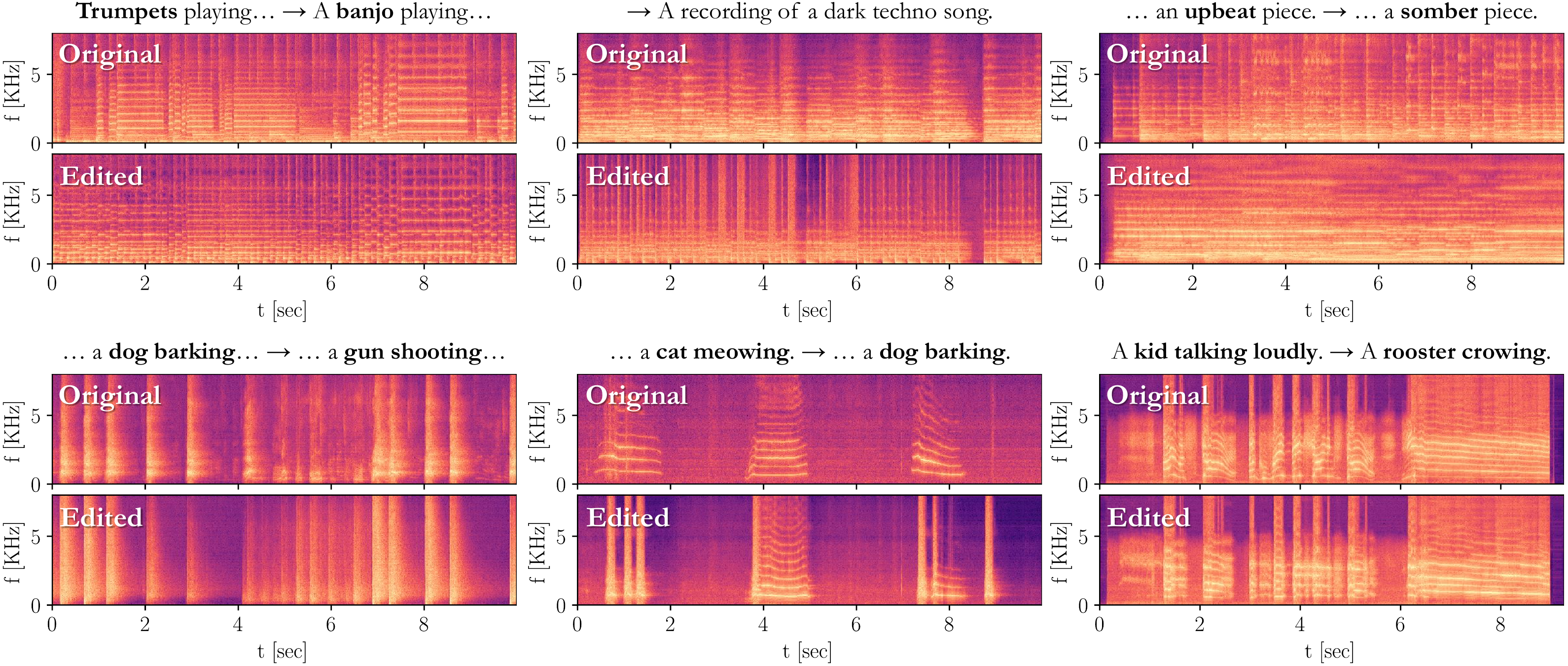}
    \caption{\textbf{Text-based zero-shot editing.} 
    ZETA editing enables changing a plethora of elements in a signal, from the genre of a song, to the objects heard in a recording.
    All examples can be listened to in \href{\urlofwebpage\#supsamples}{Sec.~1.1 of our examples page}. 
    The source and target text prompts are shown above the spectrograms, where bold marks changed text. The parameter $T_\text{start}$ (Sec.~\ref{sec:supervisedEditing}), from top to bottom and left to right, is 90, 90, 80, 100, 150, 110. The first row of original-edited pairs used the music checkpoint of AudioLDM2, while the second used the large general checkpoint.}
    \label{fig:sup_examples}
\end{figure*}

\subsection{ZETA: Text-Based Audio Editing}\label{sec:supervisedEditing}

The first editing approach we consider uses text guidance. In this setting, our goal is to edit a real audio signal $\rvx_0$ by using a text prompt $p_{\text{edit}}$ describing the desired result. Optionally, a user may also want to describe the original signal with some text prompt, $p_{\text{src}}$, so as to achieve a more fine-grained modification.

To achieve this goal, we adopt the method of \citet{huberman2023edit}, which has been previously only explored in the image domain. 
Here, we explore this approach in the context of text-to-audio models, and demonstrate its transferability to a new domain.
Specifically, we start by inverting the signal $\rvx_0$ using (\ref{eq:xt_inversion}),(\ref{eq:zt_inversion}).
We do this while injecting to the denoiser network the prompt describing the source, $p_{\text{src}}$. This is illustrated in the left pane of Fig.~\ref{fig:scheme}.
We then run the generative process (\ref{eq:reverseProcess}) with the extracted noise vectors, while injecting the prompt $p_{\text{edit}}$ describing the desired output (top-right pane of Fig.~\ref{fig:scheme}).
In both directions, we use classifier-free guidance~\citep{ho2021classifier} (CFG) for the text guidance.
The noise vectors extracted from the source signal ensure that the generated signal has the same ``coarse structure'' as the source, while the change in the text conditioning affects more fine-grained features, and leads to the editing effect. 

The balance between adhering to the target text and remaining loyal to the original signal can be controlled using two parameters. The first is the strength factor of the classifier-free guidance. Increasing this parameter steers the generation more strongly towards the desired text at the expense of departing from the original signal. The second parameter is the timestep $T_\text{start}$ from which we begin the generation process. This timestep can generally be smaller than $T$, and the smaller it is, the more the edited signal remains consistent with the source signal (see Sec.~\ref{sec:supExperiments} for examples). 

In App.~\ref{app:mask} 
we extend this editing technique to support user-provided target segments for editing. This allows more control over the resulting edited signal, which can be particularly important for editing lengthy audio data.

\subsection{ZEUS: Unsupervised Editing}\label{sec:unsupervisedEditing}

\begin{figure*}[t]
    \centering
    \begin{subfigure}{0.5\linewidth}
        \centering
        \includegraphics[width=\linewidth]{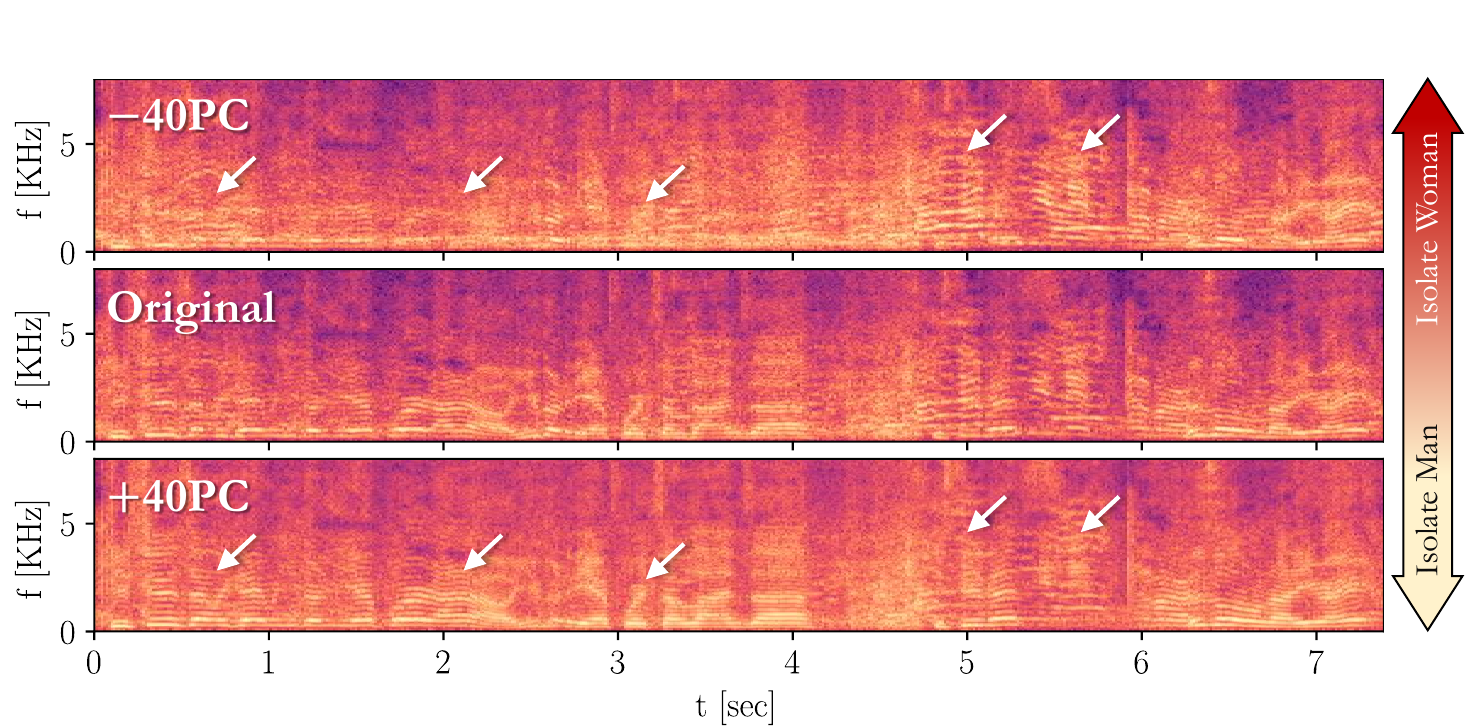}
        \caption{Unsupervised zero-shot speech editing.}
        \label{fig:unsup_speech}
    \end{subfigure}%
    \begin{subfigure}{0.5\linewidth}
        \centering
        \includegraphics[width=\linewidth]{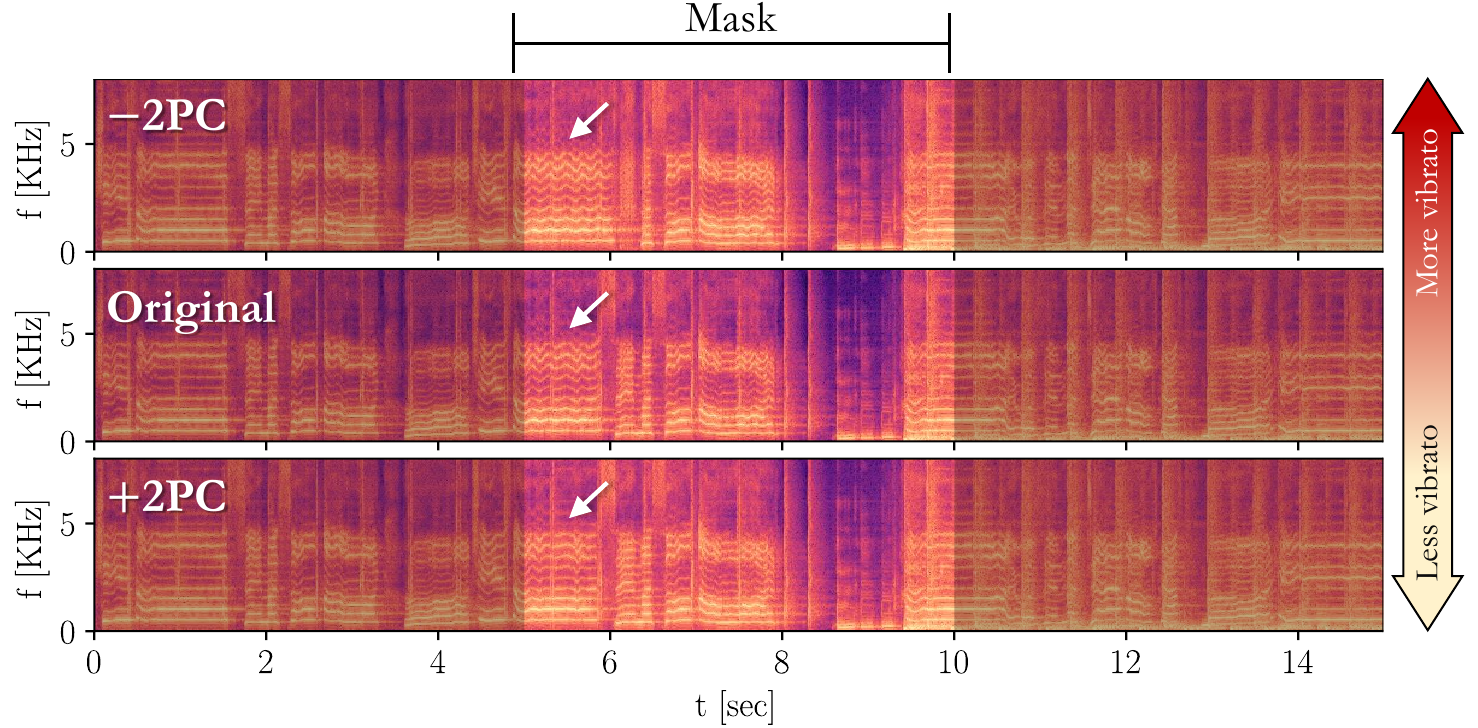}
        \caption{Unsupervised zero-shot music editing, using a mask.}
        \label{fig:unsup_mask}
    \end{subfigure}\\%
    \begin{subfigure}{\linewidth}
        \centering
        \includegraphics[width=\linewidth]{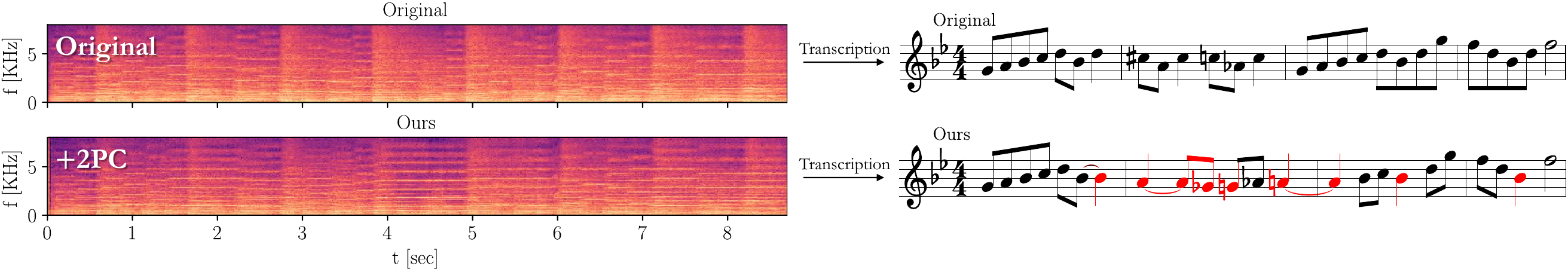}
        \caption{Unsupervised melody improvisation.}
        \label{fig:unsup_melody}
    \end{subfigure}
    \caption{\textbf{Unsupervised zero-shot editing.}
    Our ZEUS editing directions carry semantic meanings, ranging from separation of the persons in a conversation (a) to the vibrato of a singer (b) or a change in melody (c), all while retaining high semantic similarity to the source.
    Directions can be easily calculated and applied on a segment of the signal using a mask (c) (see App.~\ref{app:mask}). All examples can be listened to in \href{\urlofwebpage\#unsupsamples}{Sec.~1.2 of our examples page}.
    In (a) we use $t'=t$, $T_\text{start}=115$, $T_\text{end}=95$, in (b) we fix $t'=120$ and use $ T_\text{start}=150$, $T_\text{end}=1$, and in (c) $t'=95$, $T_\text{start}=150$, $T_\text{end}=50$. The PCs shown are the $1^{\text{st}}$ ,$3^{\text{rd}}$, and a combination of the first 3 PCs, respectively. The speech example uses the large checkpoint of AudioLDM2, without a source prompt, and the rest use the music checkpoint, with a source prompt randomly chosen from our prompts dataset.}
    \label{fig:unsup_examples}
\end{figure*}

Editing using text guidance is limited by the expressiveness of the text prompt and by the model's language understanding. 
This is arguably very significant in music, where a user may want to generate \eg variations, improvisations, or modifications to the arrangement of the piece, which are virtually impossible to precisely describe by text.
To support these kinds of edits, here we pursuit a different approach, which extracts in an unsupervised manner semantically meaningful editing directions in the noise space of the diffusion model. 
As we show, these directions can be used to perturb the generation process in multiple ways, enabling controllable semantic modifications to the signal. This does not blindly increase diversity, but rather modifies the signal in a manner that retains the essence of the original signal. Therefore, these directions can be especially effective for searching for creative inspiration or ideas given a base melody that the user composed.

As in Sec.~\ref{sec:supervisedEditing}, we start by performing edit-friendly DDPM inversion to extract noise vectors corresponding to $\rvx_0$, optionally using a text-prompt describing the signal, $p_\text{src}$. We then use those vectors in the sampling process~(\ref{eq:reverseProcess}), but with perturbations. Specifically, recall from Sec.~\ref{sec:DDPM_def} that each timestep $t$  involves $\hat{\rvx}_{0|t}$, the MSE-optimal prediction of $\rvx_0$ from $\rvx_t$. This prediction, obtained from the denoiser, corresponds to the posterior mean $\E[\rvx_0|\rvx_t]$. Our approach is to perturb this posterior mean along the top principal components (PCs) of the posterior, \ie the top eigenvectors of the posterior covariance $\Cov[\rvx_0|\rvx_t]$. This approach has been recently studied in the context of uncertainty visualization in inverse problems \citep{nehme2023uncertainty,manor2023posterior}, where it was illustrated to nicely reveal the dominant modes of uncertainty about the MSE-optimal prediction.

To compute the posterior PCs, we adapt the method proposed by \citet{manor2023posterior} to work in a generative pipeline. This work showed that the posterior covariance in Gaussian denoising is proportional to the Jacobian of the MSE-optimal denoiser.
It further showed that extracting the top eigenvectors and eigenvalues of this Jacobian can be done using the subspace iteration method~\citep{saad2011numerical}, where each iteration can be approximated using a single forward pass through the denoiser network. See App.~\ref{app:unsupEqDetails} for a detailed algorithm.

Having computed the posterior PCs $\{\rvv_{i|t'}\}$ and their corresponding eigenvectors $\{\rlambda_{i|t'}\}$ at some
timestep $t'$, we can add or subtract each of them to the denoised signal $\hat{\rvx}_{0|t}$ at every timestep $t\in[T_\text{start},\ldots,T_\text{end}]$
using a matching factor $\gamma\smash{\rlambda_{i|t}^{1/2}}$, where $\gamma$ is a user-chosen parameter controlling the strength of the modification.
As we show in App.~\ref{app:unsupEqDetails}, adding the vector $\gamma\smash{\rlambda_{i|t}^{1/2}}\rvv_{i|t'}$ to $\hat{\rvx}_{0|t}$, is equivalent to adding it to $\rvmu_t(\rvx_t)$ from Eq.~(\ref{eq:reverseProcess}) with a correction factor. Specifically, this modifies the generation process into
\begin{equation}\label{eq:newDdpmMean}
    \rvx_{t-1}= \rvmu_t(\rvx_t) + \gamma c_t\rlambda_{i|t}^{
    1/2
    }\rvv_{i|t'} + \sigma_t \rvz_t,\quad t=T,\ldots,1.
\end{equation}
where $c_t=\sqrt{\bar{\alpha}_{t-1}} - \sqrt{\bar{\alpha}_{t}}\sqrt{1-\bar{\alpha}_{t-1} -\sigma_t^2}/\sqrt{1-\bar{\alpha}_{t}}$. 
Note that instead of adding a single PC, $\rvv_{i|t'}$, we can add a linear combination of PCs, thereby creating a new direction that changes multiple semantic elements at once. 

The addition of PCs can be done in two different ways.
The first is to use $t'=t$ in (\ref{eq:newDdpmMean}), so that each denoising step is perturbed with its own PCs.
The second way involves extracting a direction from some specific timestep $t'$, and adding it at all timesteps $t\in[T_\text{start},\ldots,T_\text{end}]$, where $t'$ need not necessarily be a member of this set.
In this case, it is important to use the factor $\rlambda_{i|t}$ that matches the timestep $t$ to which the perturbation is added, and not the factor $\rlambda_{i|t'}$ corresponding to the timestep  at which the PC was computed. The role of this factor is to match the applied change to the uncertainty level at the current timestep. 

Adding the same direction at all timesteps usually strongly modifies a specific element, \eg a single note or the strength of a vibrato effect. 
Adding to each timestep its own PC can lead to a larger deviation from the original signal, \eg emphasizing a singer or changing a melody. This is because the extracted PCs can wildly differ in semantics and locality across a large range of timesteps.
Here, as in Sec.~\ref{sec:supervisedEditing}, the editing directions can be confined to a user-chosen segment by applying a mask during the computation of the PCs.

We empirically find that for each PC index $i$, the values of $\{\rlambda_{i|t}\}_{t=1}^{T}$ are similar across signals and AudioLDM2 checkpoints. We therefore compute their average value over a dataset once, and use these average values for all signals.

\section{Experiments}

To evaluate our editing methods we used AudioLDM2~\citep{liu2023audioldm2} as the pre-trained model, using 200 inference steps as recommended by the authors.
In our text-based editing experiments we compare to MusicGen~\citep{copet2023simple} conditioned on melody using their medium checkpoint, and to DDIM inversion~\citep{song2021denoising, dhariwal2021diffusion} and SDEdit~\citep{meng2021sdedit} using the same AudioLDM2 checkpoint as we use. 
DDIM inversion is typically applied for the entire diffusion process, \ie when $T_\text{start}$ is set to $T$. In App.~\ref{app:ddimmidway} we include a comparison to \emph{Partial DDIM Inversion}, a version that applies the inversion only up to $T_\text{start}$.
To evaluate our unsupervised editing method, we compare it to SDEdit, where we supply it with a prompt describing the source signal (rather than the desired edited signal). This baseline performs uncontrolled modifications to the signal, whose strength we choose using the starting timestep. 

We do not compare to AUDIT~\citep{wang2023audit} and InstructME~\citep{han2023instructme}, which train a model specifically for editing purposes, as they did not share their code and trained checkpoints. 
Additionally, we do not compare to DreamBooth and Textual Inversion as demonstrated on audio by \citet{plitsis2023investigating}, since they solve a different task -- that of personalization.
This task aims at learning a concept from a reference audio, rather than consistently modifying the input itself. Thus, personalization may allow \eg changing a genre, but cannot be used for fine-grained edits such as changing a specific instrument into a different one. In fact, as \citet{plitsis2023investigating} show, both methods have difficulty preserving the key and dynamics of the original piece, where textual inversion fails to even retain the same tempo. 
We encourage the reader to listen to our results and qualitative comparisons on our \href{\urlofwebpage}{examples page}.

Both of our methods incur a negligible memory overhead w.r.t.\ to the memory consumed by the diffusion model, and are comparable in inference speed to all evaluated competing methods. A detailed analysis can be found in App.~\ref{app:speed}.

\subsection{Datasets}

To enable a systematic analysis and quantitative comparison to other editing methods, we use the MusicDelta subset of the MedleyDB dataset~\citep{bittner2014medleydb}, comprised of 34 musical excerpts in varying styles and in lengths ranging from 20 seconds to 5 minutes, and create and release with our code base a corresponding small dataset of prompts, named \emph{MedleyMDPrompts}. This prompts dataset includes 3-4 source prompts for each signal, and 3-12 editing target prompts for each of the source prompts, totalling 107 source prompts and 696 target prompts, all labeled manually by the authors. We design the prompts to complement each other, \eg if the source prompt mentions a saxophone is playing, its target prompts may swap only the word ``saxophone'' with ``guitar'' or ``piano''. We additionally design some of the target prompts such that they do not require a complementary source prompt, and should provide enough information to edit a signal on their own (\eg for genre change). 
In our supervised text-guided experiments we randomly sub-sample a third of the source-target prompts pairs for each signal (where we include additional pairs with an empty source prompt where applicable). Thus, we evaluate our supervised prompt-based method on 324 signal-text pairs. In our unsupervised uncertainty-based experiments we randomly sub-sample one of the source-prompts per audio signal. 
Audio signals are taken from AudioSet~\citep{jort_audioset_2017}.

We remark that some works use MusicCaps~\citep{agostinelli2023musiclm} to quantitatively evaluate synthesized samples. However, this dataset contains only 10-second long music excerpts, while real music pieces can vary wildly over longer segments, changing instruments, genre or key completely. This aspect is important in the context of text-based editing, where the signal may be a minute long, and the edit should remain consistent across the entire piece (\eg when changing one instrument into another).

\begin{figure}[t]
    \centering
    \includegraphics[width=\linewidth]{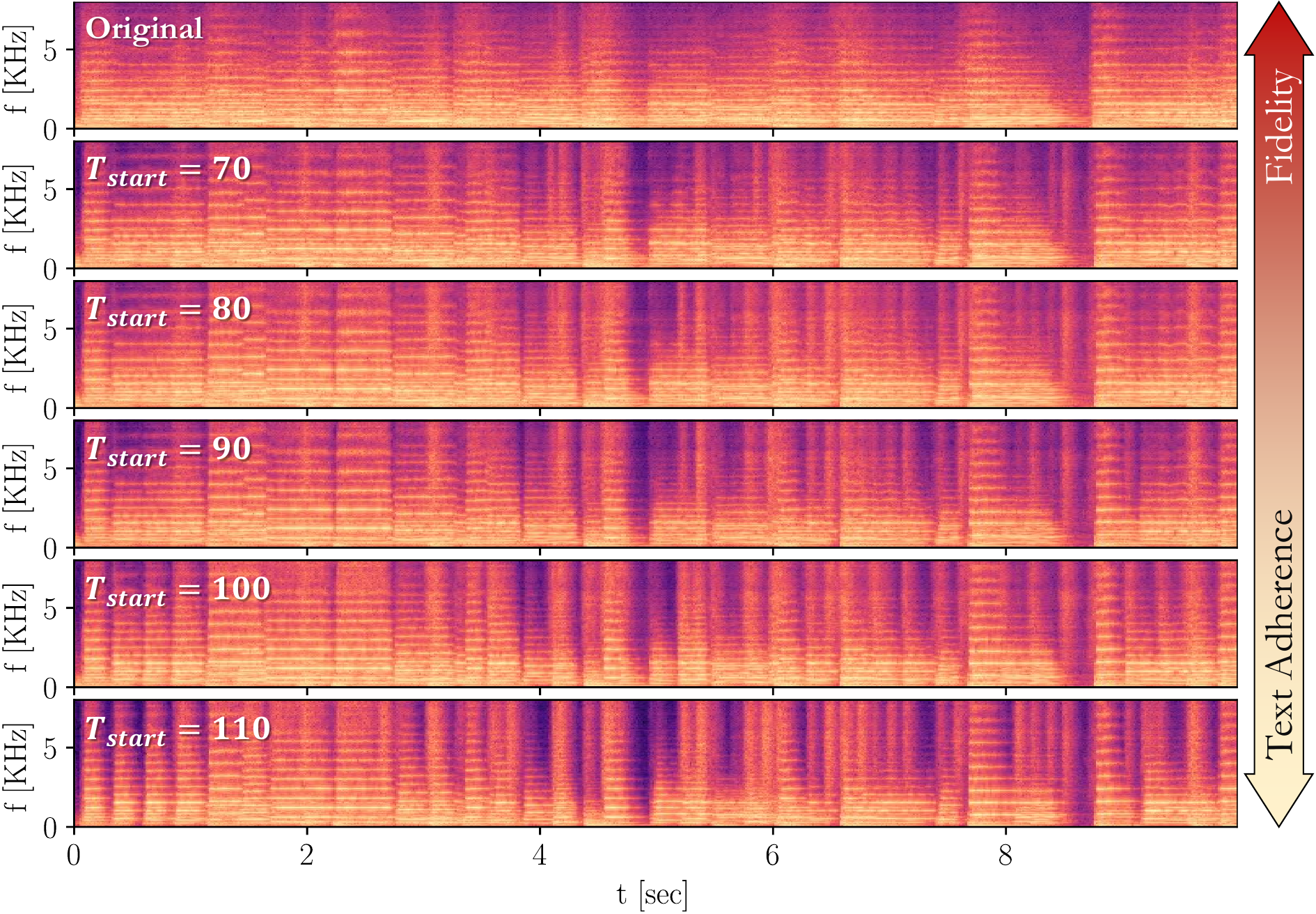}
    \caption{\textbf{The effect of $T_\text{start}$ in text-based editing.} Here, an orchestral piece is edited by using only a target prompt, $p_{\text{edit}}$=``A recording of a funky jazz song.''. 
    The signal retains more elements of the original signal as editing starts at a later timestep, denoted by $T_\text{start}$. This comes at a cost of adherence to the desired text description, \eg less coherent beats as expected by a funky jazz song, and more elongated notes as in the original orchestral piece. 
    This example can be listened to in \href{\urlofwebpagesupp\#sup-skip-effect}{Sec.~1.1.2 of our supplemental examples page}.}
    \label{fig:sup_skips}
\end{figure}

\begin{figure}[t]
    \centering
    \includegraphics[width=0.95\linewidth]{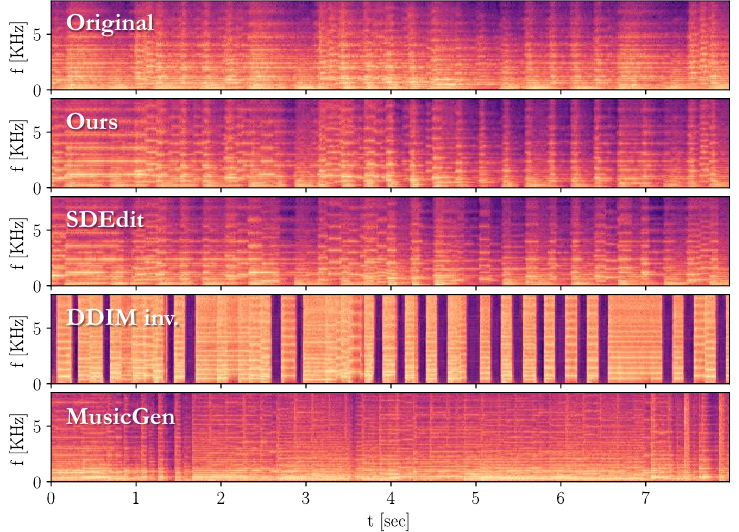}
    \caption{\textbf{Comparison of methods for text-based editing.} We compare our method, SDEdit~\citep{meng2021sdedit}, DDIM Inversion, and MusicGen~\citep{copet2023simple} for editing of the same signal. All of the results achieve a CLAP score of $\sim0.34$, however the LPAPS values are $4.29, 4.87, 6.26, 5.74$, respectively. This means that our method is most loyal to the original structure. Here for our method we use $T_\text{start}=80$ and for SDEdit $T_\text{start}=70$.}
    \label{fig:supComparison}
\end{figure}

\begin{figure*}[t]
    \centering
    \includegraphics[width=0.96\linewidth]{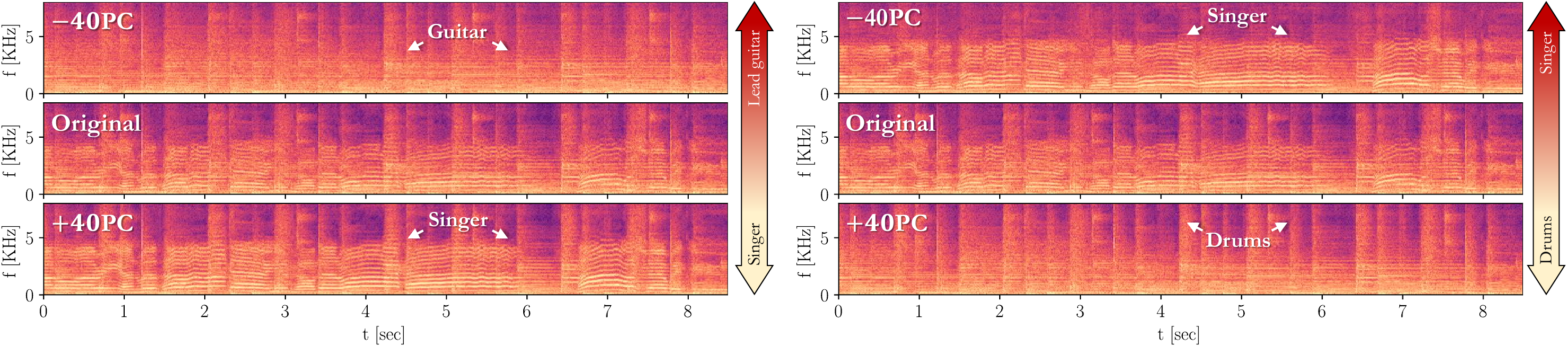}
    \caption{\textbf{The role of different PCs.} Our unsupervised editing method allows extracting different semantic editing directions for the same signal. These enable editing different semantic concepts that can complement each other. For example, the first PC shown here controls how much the singer is heard, at the expense of the lead guitar, while the second PC controls the whether the drums are dominant on the expense of the singer. Here $t'=t$, $T_{\text{start}}=115$, and $T_\text{end}=80$. This example can be listened to in \href{\urlofwebpage\#unsup-singlepc}{Sec.~1.2.2 of our examples page}.}
    \label{fig:2pcs}
\end{figure*}

\begin{figure}[t]
    \centering
    \includegraphics[width=0.8\linewidth]{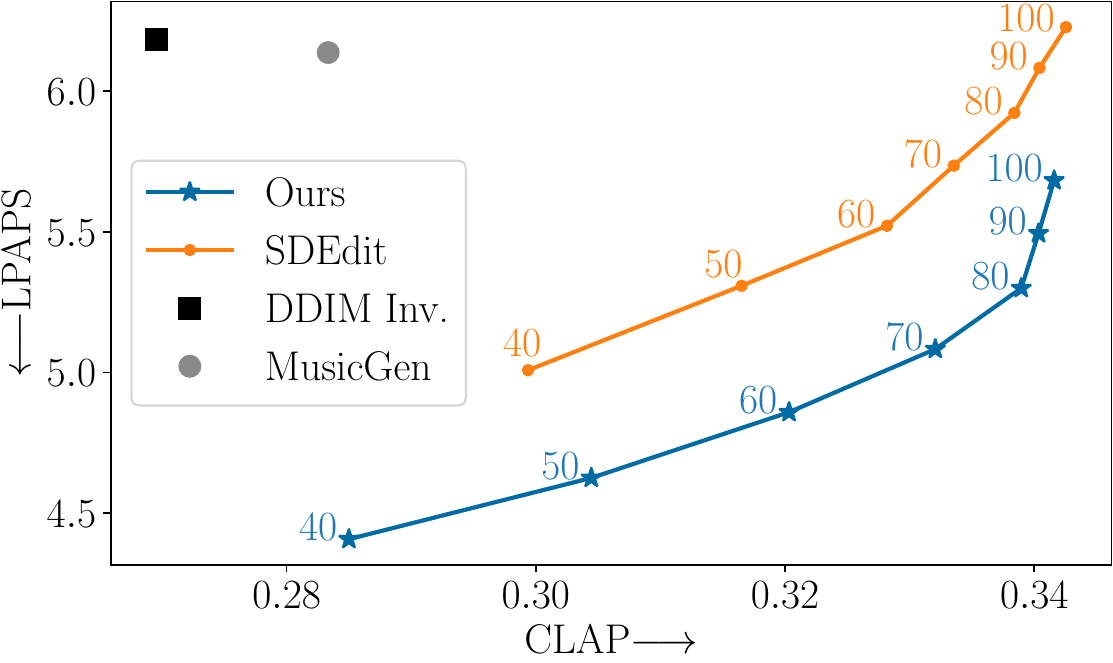}
    \caption{\textbf{Adherence to target prompt vs.~fidelity to the original signal.} The plot compares MusicGen~\citep{copet2023simple}, SDEdit~\citep{meng2021sdedit}, DDIM Inversion~\citep{song2021denoising, dhariwal2021diffusion} and our method over the MusicDelta subset in MedleyDB~\citep{bittner2014medleydb}, using our prompts dataset.
    Our method and SDEdit are shown with $T_\text{start}$ values ranging between $40$ and $100$. 
    Our results achieve lower (better) LPAPS for any level of CLAP (higher is better), indicating a good balance between text adherence and signal fidelity.} 
    \label{fig:supervised_lpapsclap}
\end{figure}

\subsection{Metrics}\label{sec:metrics}

We quantitatively evaluate the results using three types of metrics; a CLAP~\cite{laionclap2023, htsatke2022} based score to measure the adherence of the result to the target-prompt (higher is better); LPAPS~\citep{SpecVQGAN_Iashin_2021, paissan2023audio}, an audio LPIPS~\citep{zhang2018unreasonable} measure to quantify the consistency of the edited audio relative to the source audio (lower is better); and FAD~\citep{kilgour2018fr}, an audio FID~\citep{heusel2017gans} metric to measure the distance between two distributions of audio signals.
FAD has been used in the past with deep features of VGGish~\citep{hershey2017cnn} or other convolutional neural networks (CNNs) trained on VGGSound~\citep{chen2020vggsound}.
However, \citet{gui2023adapting} have shown that using such methods as a perceptual metric for music signals is sub-optimal, and so we follow their suggestion by using instead a trained large CLAP model~\citep{laionclap2023} for the deep features of the FAD calculation. 
LPAPS has also been used in the past using CNNs trained on VGGSound, nevertheless we continue with the same reasoning and use intermediate features from the same CLAP model as LPAPS' backbone in our evaluations.
In particular, we use the output layers of the four intermediate Swin-transformer blocks~\citep{liu2021swin} of the CLAP model as LPAPS' features. More details can be found in App.~\ref{app:experimentaldetails}.

\subsection{Text-Based Editing}\label{sec:supExperiments}

\begin{figure}
    \centering
    \includegraphics[width=\linewidth]{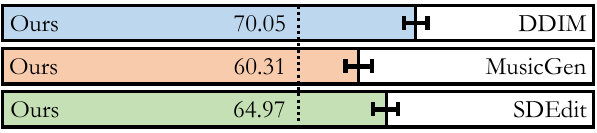}
    \caption{\textbf{User study.}
    We report the percentage of users who preferred our method over DDIM inversion~\citep{song2021denoising, dhariwal2021diffusion}, MusicGen~\citep{copet2023simple} and SDEdit~\citep{meng2021sdedit}, when answering which edited signal better matches the text prompt, while keeping the rest of the essence of the original signal.
    The error bars indicate the calculated Wilson score interval.}
    \label{fig:ustudy}
\end{figure}

Results for different effects achieved with ZETA editing are shown in Fig.~\ref{fig:teaser}(c),(d) and Fig.~\ref{fig:sup_examples}.
Fig.~\ref{fig:supComparison} shows a comparison to competing approaches. Additional comparisons can be listened to in \href{\urlofwebpage\#supcomparisons}{Sec.~2.1 of our examples page}. 
In this setting we set the CFG strength of the target prompt to $12$ for SDEdit and for our method, and to $5$ for DDIM inversion. We set this hyper-parameter such that the edits achieve a good balance between CLAP and LPIPS. The CFG strength for the source prompt is set to $3$, as recommended by \citet{liu2023audioldm}. Please see App.~\ref{app:experimentaldetails} for details on the hyper-parameters choice.
Text-based editing allows changing the global semantics of the signal, \eg by changing one instrument to another or by changing the genre of the song. 
All examples and more can be listened to in \href{\urlofwebpage\#supsamples}{Sec.~1.1 of our examples page}.

Fig.~\ref{fig:sup_skips} shows the effect of $T_\text{start}$. This parameter controls the trade-off between the adherence of the edited signal to the target prompt $p_{\text{edit}}$, and its fidelity to the original signal.
This effect can also be listened to in \href{\urlofwebpagesupp\#sup-skip-effect}{Sec.~1.1.2 of our supplemental webpage}.

To quantitatively measure the adherence of the edited signals to the target prompt and their fidelity to the source excerpts, we plot the CLAP-LPAPS results for all methods in Fig.~\ref{fig:supervised_lpapsclap}. For SDEdit and for our method we plot results for multiple $T_\text{start}$ values. 
It is evident that MusicGen, which is trained to be conditioned on a chromagram of the input signal, does not enable transferring a concept in the same way as the zero-shot editing methods we explore. As can be seen, our method outperforms all other methods under any desired balance between fidelity and text-adherence. 

\paragraph{User study.}
We additionally evaluated our method via a user study conducted through Amazon Mechanical Turk (AMT). 
AMT users were asked to answer a sequence of 16 questions, after completing a single practice question with provided feedback.
In each question, first the original music signal was played to the user, following the reveal of the target text-prompt for editing. Next, our edited result and one edit from a random competing method were played in a random order.
Users were instructed to select the edit that better matches the text prompt while keeping the rest of the essence of the original signal.
The data used for evaluation consisted of 8-second long random segments from each of the songs in the MusicDelta subset, edited with a randomly selected target text-prompt from our MedleyMDPPrompts dataset. To allow fair comparison to MusicGen, we removed prompts that referred to singing. For our method and SDEdit we arbitrarily chose $T_\text{start}=100$.
Each of the competing methods was compared using 50 workers over two batches. 
Each sequence of 16 questions included a vigilance question, during which one of the compared signals was random noise. Out of 300 participants only 277 passed the vigilance test (\ie did not choose the random noise), and their results for all methods are reported in Fig.~\ref{fig:ustudy}. As can be seen, our method was clearly preferred over all competing methods. 
More details on the user study can be found in App.~\ref{app:userStudy}.

\subsection{Unsupervised Editing}\label{sec:unsupExperiments}

\begin{figure}[t]
    \centering
    \includegraphics[width=0.8\linewidth]{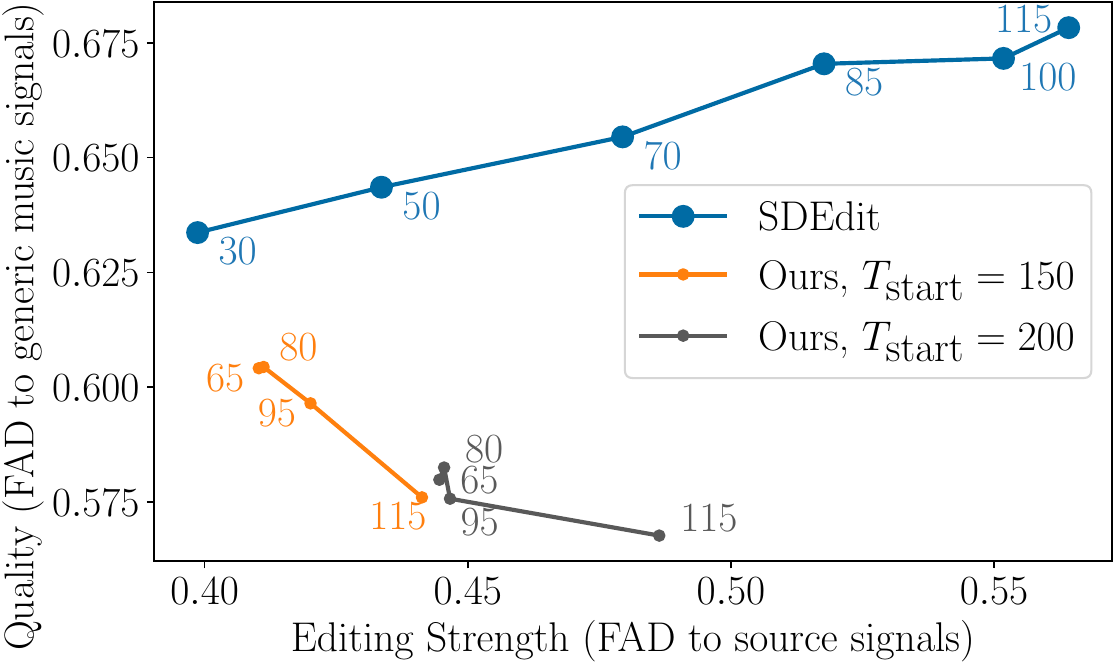}
    \caption{\textbf{Unsupervised editing strength vs.~quality.} We compare SDEdit~\citep{meng2021sdedit} to our method over the MusicDelta subset of MedleyDB~\citep{bittner2014medleydb}, using our prompts dataset.
    SDEdit is shown with $T_\text{start}$ values ranging between $30$ and $115$, and our method is shown for different $t'$ values (indicated on the plot) and $T_\text{start}$ values (indicated in the legend). We use $T_\text{end}=1$. For any level of perceptual deviation from the original signal, our method retains a higher quality (w.r.t.~FMA-pop~\citep{gui2023adapting,fma_dataset}).} 
    \label{fig:unsup_fadfad}
\end{figure}

Next, we perform experiments using ZEUS editing. 
Here we set the CFG strength to $3$ for both SDEdit and our method. 
The strength of the modification is controlled by the parameter $\gamma$. As can be seen in Figs.~\ref{fig:teaser}(a),(b),~\ref{fig:unsup_examples}, and \ref{fig:2pcs}, and can be listened to in \href{\urlofwebpage\#unsupsamples}{Sec.~1.2 of our examples page}, the modifications can range from effects like voice emphasis, to more stylistic changes \eg in a singer's vibrato. An interesting change is more apparent in music, where the semantic editing directions can take the form of an improvisation on the original piece, obtained by changing the melody, as shown in Fig.~\ref{fig:unsup_melody}. 

The PCs of the posterior covariance convey the uncertainty of the denoising model at the current timestep. The synthesis process is inherently more uncertain at earlier timesteps in the sampling process (\ie at larger $t$). Therefore, the extracted directions $\{\rvv_{i|t}\}$ generally exhibit more global changes spread over larger segments of the samples for earlier timesteps, and more local changes for later timesteps.
Empirically, above a certain timestep the extracted directions are not interesting. We  therefore restrict ourselves here to $t\le135$ (see App.~\ref{app:UnsupEntropy} for further discussion).

Qualitative comparisons can be listened to in \href{\urlofwebpage\#unsupcomparisons}{Sec.~2.2 of our examples page}.
Quantifying unsupervised edits in music can be challenging as the editable aspects of music vary widely, from rhythm and melody to instrumentation, mood and genre. Therefore, for this task, LPAPS is less fitting for measuring the ``strength'' of an edit. Specifically, a semantically small change like a slight shift in rhythm that occurs across the entire signal can throw off LPAPS completely, even though it is barely perceived by humans. Similarly, a short melodic change will achieve a small LPAPS distance, but can significantly shift the perceived mood of the piece. 

Therefore, instead of LPAPS, here we measure the FAD to two different datasets. 
The first is the original Music Delta subset. This measures the strength of the edit, as it quantifies the deviation from the original distribution. The second is the FMA-pop dataset, a subset of FMA~\citep{fma_dataset} proposed by \citet{gui2023adapting}. This subset contains the 30 most popular songs for each of the 163 genres in the FMA dataset, and as such contains a large variety of genres and styles. 
This FAD gives an idea about the musical quality of the edited output on its own. This is because the songs in FMA-pop corresponds to the top listens and thus represent ``likeability''.
We plot the two metrics in Fig.~\ref{fig:unsup_fadfad}, using different $T_\text{start}$ configurations for SDEdit and different $t$ and $t'$ values for our approach.
Our method achieves a higher quality (lower FAD to general music) for any desired edit strength. 
See App.~\ref{app:randVector} for validation of the semantics of our directions, and App.~\ref{app:images} for their applicability in other domains, \eg images.

\section{Conclusion}

We presented two methods for zero-shot editing of audio signals using pre-trained diffusion models. To the best of our knowledge, this is the first attempt to fully explore zero-shot editing in the audio domain. 
In addition to a text-based method, which we adopted from the image domain, we proposed a novel unsupervised method for discovering editing directions. 
As in all zero-shot editing methods, the quality of the edited outputs are limited by the quality of the pre-trained audio model. This is most noticeable in text-based editing, as lack of proficiency in the textual information given in the target prompt can limit the edited output.
In unsupervised editing this concern is mitigated since no prompt is needed. However, as in all unsupervised methods, there rises the drawback of uncontrollability of the extracted PCs. A user cannot know in advance their PC of interest, and finding a PC for a user's liking is not guaranteed. 
Despite these limitations, we demonstrated both qualitatively and quantitatively that our methods outperform other methods for text-based editing, and illustrated that our unsupervised method is able to create semantically meaningful modifications and improvisations to a source signal.

\section*{Impact Statement}

The purpose of this paper is to advance the field of Machine Learning and in particular zero-shot editing of audio signals.
We feel that there are many potential societal consequences of our work, but the predominant one relates to the ability of using our method for copyright infringement.
In this work we worked on audio licensed under Creative Commons Attribution, and as this is an academic work it is in fair use.
However, users might use our methods to modify existing copyrighted musical pieces without sufficient permission of the copyright holder, and this might not fall under fair use under different circumstances.
We believe it is important to develop methods for automatically detecting whether AI-based methods have been applied to audio signals.

\section*{Acknowledgements}
This research was partially supported by the Israel Science Foundation (grant no. 2318/22), by a gift from Elbit Systems, and by the Ollendorff Minerva Center, ECE faculty, Technion.

\bibliography{citations}
\bibliographystyle{icml2024}

\newpage
\appendix

\onecolumn


\section{Partial DDIM Inversion}\label{app:ddimmidway}

DDIM inversion~\cite{song2021denoising,dhariwal2021diffusion} is usually used only when $T_\text{start}$ is set to $T$.
The intuition behind this plain manner is that in DDIM sampling, the denoising process is deterministic given the noise map at $t=T$.
The DDIM inversion process extracts this single noise map at $t=T$, and is usually not stopped at some mid-way timestep. 
The disadvantage of this approach is that it accumulates error which leads to poor reconstruction even when using the same prompt for the generation stage~\citep{mokady2023null}. 
Contrary to that, DDPM sampling is a stochastic process, and therefore stopping DDPM inversion at any $t$ is both theoretically motivated and can provide different edited results for the same text-prompt, using different inverted noise maps. 
The choice of $t$ for DDPM inversion is completely up to the user and allows for determining how much the edited signal should diverge from the original signal. 

\begin{figure}[t]
    \centering
    \includegraphics[width=0.8\linewidth]{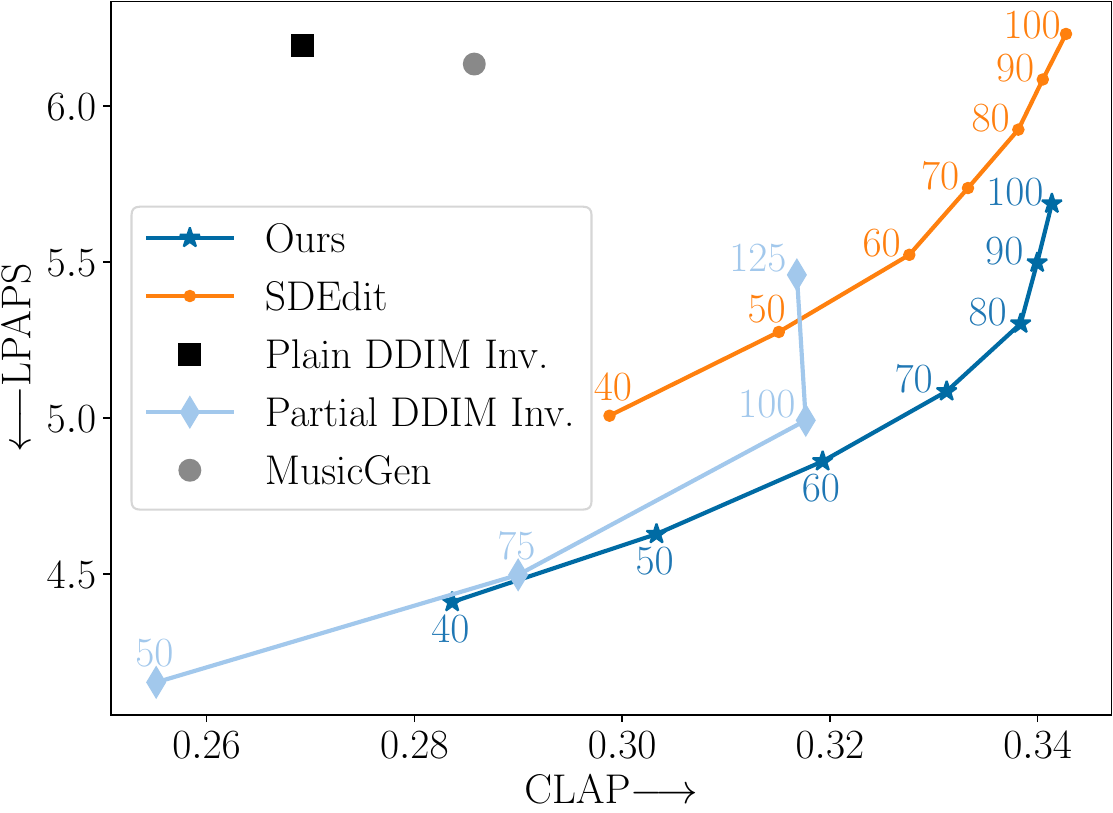}
    \caption{\textbf{Adherence to target prompt vs.~fidelity to the original signal.} 
    The plot compares MusicGen~\citep{copet2023simple}, SDEdit~\citep{meng2021sdedit}, DDIM Inversion~\citep{song2021denoising, dhariwal2021diffusion} and our method over the MusicDelta subset in MedleyDB~\citep{bittner2014medleydb}, using our prompts dataset.
    Our method and SDEdit are shown with $T_\text{start}$ values ranging between $40$ and $100$. 
    DDIM Inversion is shown both in its plain version, and its partial version when stopped mid-way using $T_\text{start}$ values ranging between $50$ and $125$.
    Our results achieve lower (better) LPAPS for any level of CLAP (higher is better), indicating a good balance between text adherence and signal fidelity.} 
    \label{fig:supervised_lpapsclap_ddim}
\end{figure}

Nevertheless, for the sake of completeness we choose to also compare here to a version of DDIM inversion that does stop mid-way, hence the name \emph{Partial DDIM Inversion}. We use the same CFG parameters used for the plain DDIM Inversion approach, and plot the CLAP-LPAPS results for our method, the competing methods, and partial DDIM inversion for multiple $T_\text{start}$ values, in the same manner as in Sec.~\ref{sec:supExperiments}. 
For each method the $T_\text{start}$ values range is shown up to its turning point, that is due to the divergence between the training dataset distributions of CLAP and the text-encoder of AudioLDM2.
The results appear in Fig.~\ref{fig:supervised_lpapsclap_ddim}. 
Interestingly, this uncommon yet simple approach achieves relatively good results, however it is a bit lacking in text-adherence compared to other methods.

\section{Experimental Details}\label{app:experimentaldetails}

For the CLAP model used in the CLAP, LPAPS, and FAD metrics calculation, as described in Sec.~\ref{sec:metrics}, we follow \citet{gui2023adapting} and MusicGen~\citep{copet2023simple}, and use the ``music\_audioset\_epoch\_15\_esc\_90.14.pt'' checkpoint of LAION-AI~\citep{htsatke2022,laionclap2023}.
Since this checkpoint was trained for 10-second long segments, to calculate the score of a signal we first split the signal into overlapping 10-second long segments, calculate the score of each segment separately, and take the mean of the scores as the score of the entire signal.
In all evaluations and for both metrics we use an overlap of one second.

In all of our unsupervised editing experiments, we run 50 subspace iterations for extracting PCs, and set $C=10^{-3}$ as the small approximation constant as described by \citet{manor2023posterior}. 
We use MusicGen with their default parameters provided in their official implementation demo. Additionally, we opt to not use negative prompts in all experiments.

\begin{figure}
    \centering
    \includegraphics[width=0.7\linewidth]{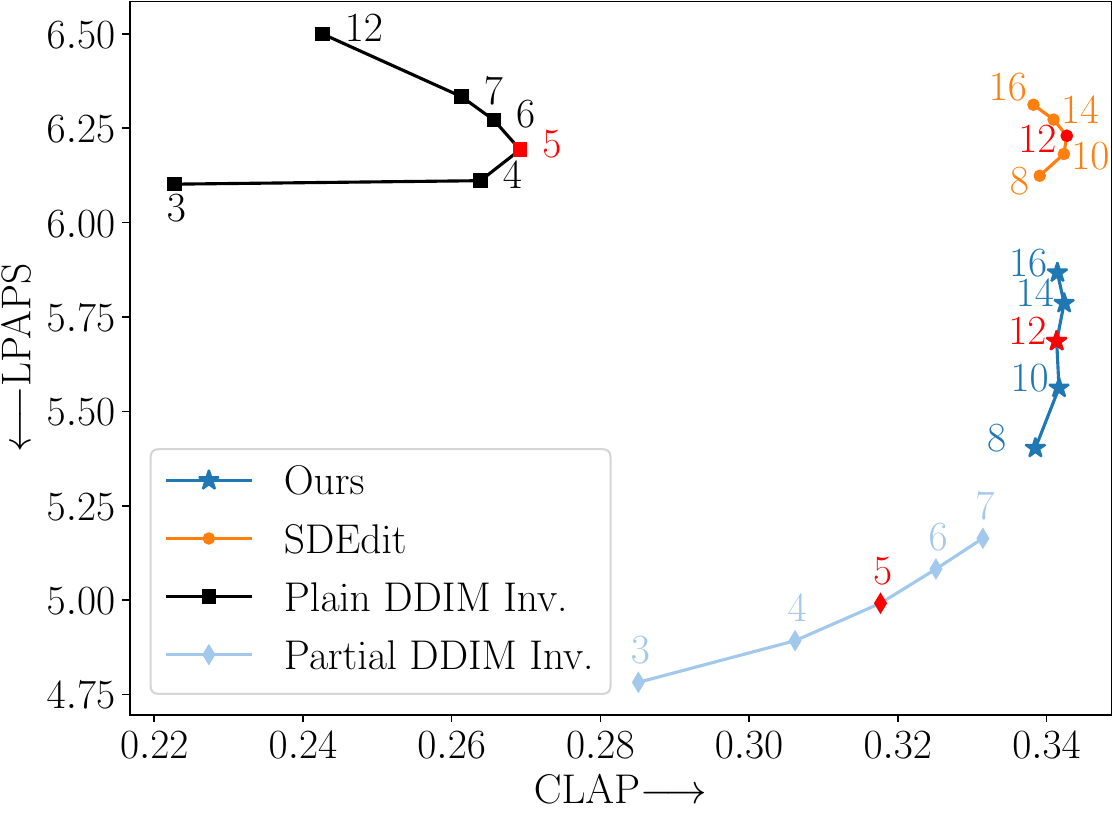}
    \caption{
    \textbf{Comparison of different target classifier-free guidance strengths used in the text-based editing process.}
    The plot compares SDEdit~\cite{meng2021sdedit}, DDIM Inversion~\cite{song2021denoising, dhariwal2021diffusion} and our method over the MusicDelta subset in MedleyDB~\citep{bittner2014medleydb}, using our prompts dataset.
    Our method and SDEdit are shown with target-prompt CFG strengths ranging between 8 and 16, when $T_\text{start}$ is set to $100$. 
    DDIM Inversion is shown both in its plain version ($T_\text{start}=200$), with target-prompt CFG strengths ranging between 3 and 12, and in its partial version when $T_\text{start}$ is set to $100$, with target-prompt CFG strengths ranging between 3 and 7. The floating numbers indicate the target-prompt CFG strength for each method.
    The chosen strength for SDEdit (12) and DDIM Inversion (5) are the best considering the trade-off between text-adherence and fidelity to the original signal. Our method can achieve even slightly better LPAPS scores while retaining the same CLAP score when lowering the strength below the value of 12, which we used in our experiments.}
    \label{fig:target_cfg}
\end{figure}

\begin{figure}
    \centering
    \includegraphics[width=0.7\linewidth]{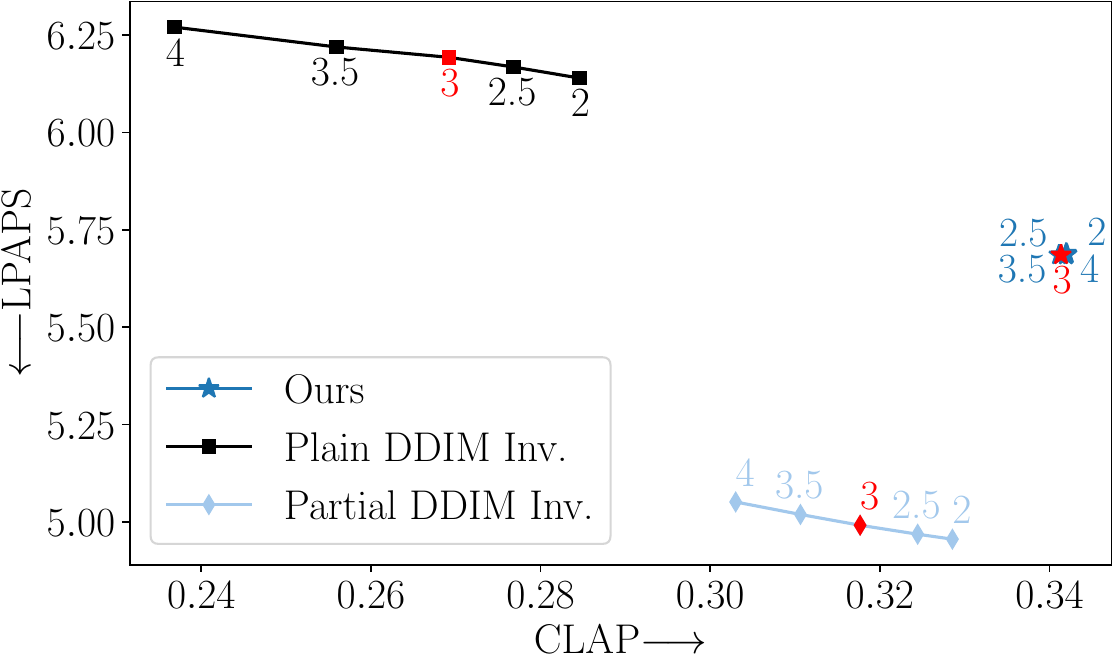}
    \caption{
    \textbf{Comparison of different source classifier-free guidance strengths used in the text-based editing process.}
    The plot compares DDIM Inversion~\cite{song2021denoising, dhariwal2021diffusion} and our method over the MusicDelta subset in MedleyDB~\citep{bittner2014medleydb}, using our prompts dataset.
    All methods are shown with source-prompt CFG strengths ranging between 2 and 4. DDIM version is shown both in its plain version ($T_\text{start}=200$, and in its partial version. For both our method and partial DDIM inversion $T_\text{start}$ is set to $100$.
    The floating numbers indicate the source-prompt CFG strength for each method.
    The chosen strength for DDIM inversion (3) is a balanced choice, whereas for our method different strengths achieve largely similar scores.}
    \label{fig:source_cfg}
\end{figure}

\paragraph{Classifier-Free Guidance strength.}
The classifier-free guidance strength for the source parameter was set to 3, as recommended by \citet{liu2023audioldm}. In the unsupervised editing approach, this is the only classifier-free guidance strength hyper-parameter, as the prompt (if used) and its strength is not changed during the editing process.
In the text-based editing approach, the strength used for the target prompt was chosen such that the resulting edits achieve a good balance between their CLAP and their LPAPS scores.
To verify the chosen classifier-free guidance strengths for the different methods, we conduct two ablations, and display the results in Fig.~\ref{fig:target_cfg} and Fig.~\ref{fig:source_cfg}.
We include in the ablation the results for the partial DDIM inversion approach discussed in App.~\ref{app:ddimmidway}, and set its $T_\text{start}$ value to $100$. We set the same $T_\text{start}$ value for our method for and SDEdit~\cite{meng2021sdedit}.
For the target CFG ablation, we keep the source prompt guidance strength set to 3 for all methods. 
The results depicted in Fig.~\ref{fig:target_cfg} show that the chosen strength for SDEdit (12) and DDIM Inversion~\citep{song2021denoising, dhariwal2021diffusion} (5) are the best considering the trade-off between text-adherence, measured with CLAP, and fidelity to the original signal, measured with LPAPS. Our method can achieve even slightly better LPAPS scores when retaining the same CLAP score when lowering the strength below the value of 12, which we used in our experiments.
In the source CFG ablation, we keep the chosen target prompt guidance strengths, \ie 5 for DDIM inversion and 12 for our method. 
As can be seen in Fig.~\ref{fig:source_cfg}, the chosen hyper-parameter is a balanced choice for DDIM inversion, whereas for our method the different strengths achieve simliar CLAP and LPAPS scores.

\section{Memory Requirements and Inference Speed}\label{app:speed}

Let $T$ be the total number of sampling steps (\eg in all our experiments we used $T=200$), and $S$ a Boolean indicating whether a source prompt is used. 
Whenever any prompt is used by applying classifier-free guidance, this involves two passes through the diffusion model's UNet, and therefore two neural function evaluations (NFEs).

We will start by addressing the text-based editing approach.
Our method and DDIM Inversion require $(3+S)T_{start}$ neural function evaluations (NFEs). $(1+S)T_{start}$ NFEs account for the inversion process (which runs from $t=0$ up to $t=T_{start}$), and $2T_{start}$ NFEs account for the prompt-accompanied generation (which runs from $t=T_{start}$ to $t=0$).
For DDIM inversion, $T_{start}$ is usually always taken to be $T$, as discussed in App.~\ref{app:experimentaldetails}. We also compare to a DDIM inversion process which stops mid-way, so this DDIM inversion version takes the same amount of NFEs as our method.   
SDEdit adds noise to the signal and then generates the edited output using $2T_{start}$ NFEs. Note, however, that SDEdit does not allow using a source prompt to help guide the editing process.
MusicGen is a non-diffusion based model, trained specifically for editing, and is therefore not comparable using NFEs.
From our experiments, for a 30 seconds long audio signal, a single NFE takes an average 123 milliseconds to complete. In a realistic editing scenario, without a source prompt (\ie setting $S=0$), and setting $D=200, T_{start}=100$, our method takes 37 seconds to complete, while MusicGen takes 35.5 seconds.
All methods except MusicGen can easily set a lower $D$ to shorten the sampling time substantially.

In the unsupervised editing approach, we will assume for this calculation no prompts. When a prompt is used to accompany the editing process, the results are just multiplied by a factor of two.
Here we have an additional overhead in the calculation of the PC, which requires running $K$ subspace iterations (per PC), meaning $K$ NFEs.
In our experiments we set $K=50$, resulting in a 6-seconds long overhead for a 30-seconds long signal. We note that empirically we saw that a smaller $K$ can suffice as well.
In the unsupervised case, SDEdit still takes $T_{start}$ NFEs. For our method, when $t'\\le T_{start}$, we first invert the signal up to $T_{start}$, then compute the PCs, and finally run the generation, totaling in $2T_{start}+K$ NFEs. When $t'>T_{start}$, we extend the inversion process up to $t'$, yielding $t'+T_{start}+K$ NFEs.

Finally, we consider the memory consumption of our methods. In all methods, both the used diffusion model and the noise tensors must be held in the memory of the GPU. 
We find that in the average case, the memory overhead of the noise tensors are negligible w.r.t.\ to the memory consumed by the diffusion model itself. For example, for a 30-seconds long signal, 200 noise tensors occupy 0.65GB, compared to 5.6GB occupied by the diffusion model.
\section{Editing Over a User-Chosen Segment}\label{app:mask}

Editing can be confined to a user-chosen segment, rather than the whole signal, by using a mask during the generative process.
When doing so, at each timestep $t$, after computing $\rvx_{t-1}$ using (\ref{eq:reverseProcess}) or (\ref{eq:newDdpmMean}), for text-based editing and unsupervised editing respectively, we enforce the parts of the signal outside the mask to shift back closer to the original signal at that timestep. We do this by setting
\begin{equation}
    \rvx_{t-1} \leftarrow M\odot\rvx_{t-1} + (1-M)\odot(\delta\rvx^{\text{orig}}_{t-1} + (1-\delta)\rvx_{t-1}),
\end{equation}
where $M$ is the mask, and $\delta$ is some small constant which we fix to $0.025$ in all experiments.

\begin{figure}[t]
    \centering
    \includegraphics[width=\linewidth]{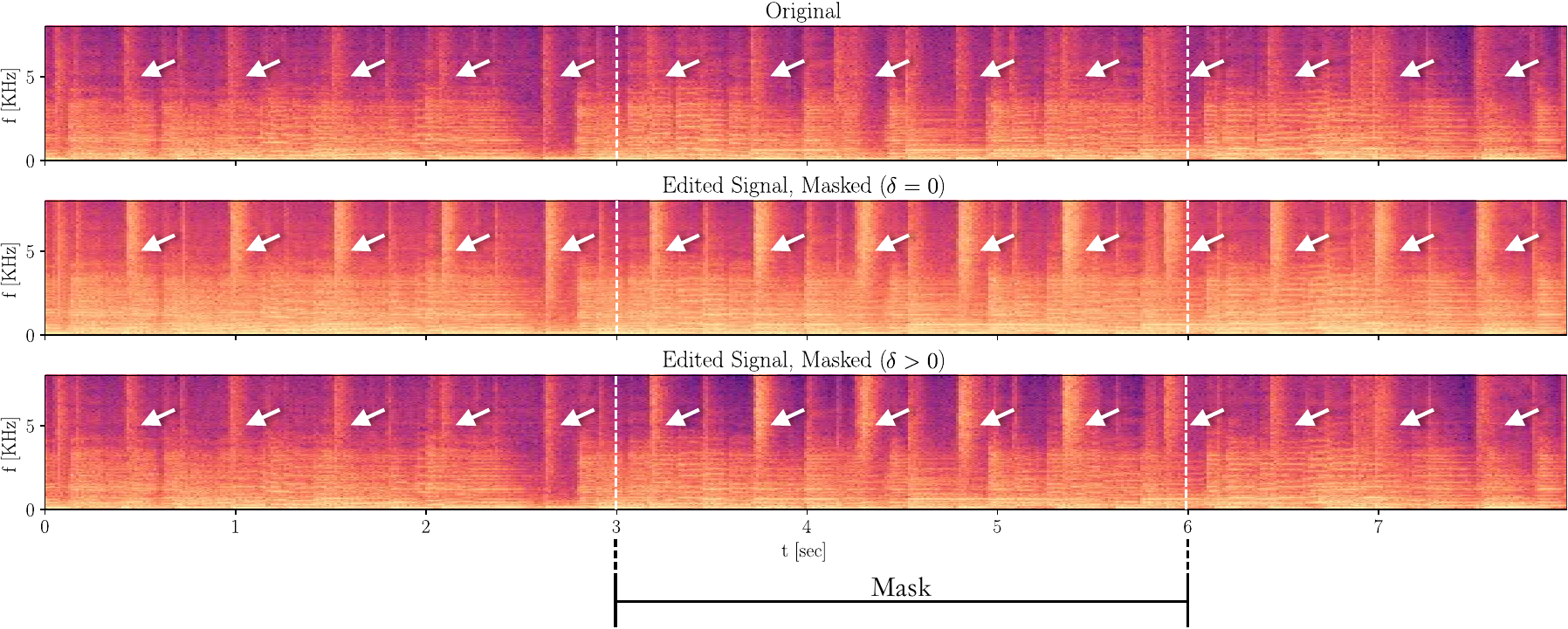}
    \caption{\textbf{Example of data leakage in diffusion models editing.} The extracted PC affects the snare of the drums beats, marked with arrows. By shifting the parts of the signal outside the mask closer to the original signal at each timestep, the snare only changes in the masked region. By not shifting the signal, the snare changes along the entire signal.}
    \label{fig:maskfix}
\end{figure}

This shift is necessary due to ``data leakage'' caused by the architecture of the diffusion model.
The term data leakage refers to the phenomenon where a localized edit, \eg a PC calculated for a specific segment of the signal using a mask, unintentionally affects the rest of the signal.
This is commonly caused by the use of a UNet and attention modules as backbones for the diffusion model. 
The effect of the data leakage can be viewed in Fig.~\ref{fig:maskfix}.
The extracted PC at this timestep, $t=80$, affects the snare of the drums beats. 
Without shifting the parts outside the mask back to the original signal, the snare changes across the entire signal. By setting $\delta>0$ the change is localized to the masked region.

We notice empirically that the strength of the data leakage depends on the type of edit, however, generally it is not known a priori. 
Additionally, we would like to note that setting $\delta>0$ is application specific and its effect is subjective. 
On the one hand, using a mask implies that changes outside the mask region are unwanted.
However, allowing the edit to have a more global influence across the entire signal, could result in a more consistent result (\eg it might make more sense that the drum beats' snare changes across an entire musical piece). 

\section{Unsupervised Editing Implementation Details}\label{app:unsupEqDetails}

Consider the multivariate denoising problem observation $\rvy=\rvx+\rvn$, where $\rvx$ is a random vector and the noise $\rvn\sim \gN(0,\sigma^2)$ is statistically independent of $\rvx$. Then, \citet{manor2023posterior} show that the posterior covariance relates to the Jacobian of an MSE-optimal denoiser,
\begin{equation}
    \Cov[\rvx|\rvy=\vy]= \sigma^2\frac{\partial\E[\rvx|\rvy=\vy]}{\partial\vy},
\end{equation}
where $\vy$ is the noisy sample. The work further showed that extracting the top eigenvectors and eigenvalues of the posterior covariance can be done using the subspace iteration method~\citep{saad2011numerical}, where each iteration can be approximated using a single forward pass through the denoiser network.

In DDMs, Eq.~(\ref{eq:forwardProcess}) can be rearranged to fit the aforementioned observation model:
\begin{equation}
    \frac{\rvx_t}{\sqrt{\bar{\alpha}_t}} = \rvx_0 + \frac{\sqrt{1-\bar{\alpha}_t}}{\sqrt{\bar{\alpha}_t}}\rvepsilon_t, \quad t=1,\ldots,T.
\end{equation}
Therefore, $\rvx=\rvx_0$, $\rvy=\rvx_t/\sqrt{\bar{\alpha}_t}$, and $\rvn\sim\gN(0,(1-\bar{\alpha}_t)/\bar{\alpha}_t\mI)$.
Under this point of view, each timestep $t$ in the diffusion process is a Gaussian denoising problem, and we get 
\begin{align}
    \Cov\left[\rvx_0\middle|\frac{\rvx_t}{\sqrt{\bar{\alpha}_t}}\right] &= \frac{1-\bar{\alpha}_t}{\bar{\alpha}_t} \cdot \frac{\partial  \E\left[\rvx_0\middle|\frac{\rvx_t}{\sqrt{\bar{\alpha}_t}}\right]}{\partial\frac{\rvx_t}{\sqrt{\bar{\alpha}_t}}} \nonumber \\
    \Cov\left[\rvx_0\middle|\rvx_t\right] &= \frac{1-\bar{\alpha}_t}{\bar{\alpha}_t} \cdot \frac{\partial  \E\left[\rvx_0\middle|\rvx_t\right]}{\partial\rvx_t}.
\end{align}
This allows for using the algorithm proposed by \citet{manor2023posterior} at every desired timestep $t'$ of the diffusion model, to extract semantic directions $\{\rvv_{i|t'}\}$ and their corresponding factors $\{\rlambda_{i|t'}\}$.

These directions are PCs of the posterior covariance, and as such they need to be added to the primal space, and in particular to $\hat{\rvx}_{0|t'}$.
As described in Sec.~\ref{sec:DDPM_def}, the reverse process of a diffusion model is written in Eq.~(\ref{eq:reverseProcess}) as $\rvx_{t-1}= \rvmu_t(\rvx_t) + \sigma_t \rvz_t$, where $\{\rvz_t\}_{t=1}^T\sim\gN(0,\mI)$, $\rvz_0=0$, $\{\sigma_t\}$ is an increasing sequence of noise levels, and 
$\rvmu_t(\rvx_t)$ is a function of a neural network trained to predict $\rvepsilon_t$ from $\rvx_t$.

This function can be expressed as the sum of two elements:
\begin{equation}\label{eq:ddpmMean}
    \rvmu_{t}= \sqrt{\bar{\alpha}_{t-1}} \mP\big(\rvf_t(\rvx_t)\big) + \mD\big(\rvf_t(\rvx_t)\big).
\end{equation}
Here, $\rvf_t(\rvx_t)$ is the aforementioned neural network trained to predict $\rvepsilon_t$ from $\rvx_t$. $\mP\big(\rvf_t(\rvx_t)\big)$ is given by 
\begin{equation}\label{eq:Pdef}
    \mP\big(\rvf_t(\rvx_t)\big)= \big(\rvx_t-\sqrt{1-\bar{\alpha}_t}\rvf_t(\rvx_t)\big)/\sqrt{\bar{\alpha}_t},
\end{equation}
and is the predicted $\rvx_0$ at timestep $t$, noted as $\hat{\rvx}_{0|t}$. $\mD\big(\rvf_t(\rvx_t)\big)$ is the direction pointing to $\rvx_t$, given by
\begin{equation}\label{eq:Ddef}
    \mD\big(\rvf_t(\rvx_t)\big)=\sqrt{1-\bar{\alpha}_{t-1} -\sigma_t^2}\rvf_t(\rvx_t).
\end{equation}
Note that $\mP\big(\rvf_t(\rvx_t)\big)$ gives a simple connection between $\hat{\rvx}_{0|t}$ and $\rvf_t(\rvx_t)$, the neural network, and therefore changing one is equivalent to changing another. Specifically, suppose we apply an edit to $\rvmu_{t}$ by adding $\gamma\sqrt{\rlambda_{i|t}}\rvv_{i|t'}$ to $\hat{\rvx}_{0|t}$ in Eq. (\ref{eq:ddpmMean}).
We denote this edited $\rvmu_{t}$ as $\rvmu_{t}^\text{edit}$:
\begin{equation}\label{eq:asym_apply_vt_inx0}
    \rvmu_{t}^\text{edit}(\rvx_t) = \sqrt{\bar{\alpha}_{t-1}} \Big(\mP\big(\rvf_{t}(\rvx_{t})\big) + \gamma \sqrt{\rlambda_{i|t}} \rvv_{i|t'}\Big) + \mD\big(\rvf_{t}(\rvx_{t})\big).
\end{equation}
By substituting Eq.~(\ref{eq:Pdef}) into Eq.~(\ref{eq:asym_apply_vt_inx0}) we get
\begin{align}
    \rvmu_{t}^\text{edit}(\rvx_t) &= \sqrt{\bar{\alpha}_{t-1}} \Big(
    \frac{\rvx_t-\sqrt{1-\bar{\alpha}_t}\rvf_t(\rvx_t)}{\sqrt{\bar{\alpha}_t}}
    + \gamma \sqrt{\rlambda_{i|t}} \rvv_{i|t'}\Big) + \mD\big(\rvf_{t}(\rvx_{t})\big) \nonumber \\
    &= \sqrt{\bar{\alpha}_{t-1}} \cdot \frac{\rvx_t-\sqrt{1-\bar{\alpha}_t}\rvf_t(\rvx_t) + 
    \gamma \sqrt{\bar{\alpha}_t} \sqrt{\rlambda_{i|t}} \rvv_{i|t'} }{\sqrt{\bar{\alpha}_t}}
    + \mD\big(\rvf_{t}(\rvx_{t})\big) \nonumber \\
    &= \sqrt{\bar{\alpha}_{t-1}} \cdot \frac{\rvx_t-\sqrt{1-\bar{\alpha}_t} \big( \rvf_t(\rvx_t) - \gamma \frac{\sqrt{\bar{\alpha}_t}}{\sqrt{1-\bar{\alpha}_t}} \sqrt{\rlambda_{i|t}} \rvv_{i|t'} \big)}{\sqrt{\bar{\alpha}_t}}
    + \mD\big(\rvf_{t}(\rvx_{t})\big) \nonumber \\
    &= \sqrt{\bar{\alpha}_{t-1}} \mP\Big(\rvf_{t}(\rvx_{t}) -  \gamma \frac{\sqrt{\bar{\alpha}_t}}{\sqrt{1-\bar{\alpha}_t}} \sqrt{\rlambda_{i|t}} \rvv_{i|t'} \Big)
    + \mD\big(\rvf_{t}(\rvx_{t})\big).
\end{align}
This is an asymmetric reverse process formulation (Asyrp), since the two functions $\mP$ and $\mD$ that compose $\rvmu_t^\text{edit}$ are operating on different variables.
A symmetric reverse process is then given by also changing $\mD\big(\rvf_t(\rvx_t)\big)$ accordingly,
\begin{equation}\label{eq:sym_apply_vt_inx0}
    \rvmu_{t}^\text{edit}(\rvx_t) = \sqrt{\bar{\alpha}_{t-1}} \mP\Big(\rvf_{t}(\rvx_{t}) - \gamma \frac{\sqrt{\bar{\alpha}_t}}{\sqrt{1-\bar{\alpha}_t}} \sqrt{\rlambda_{i|t}} \rvv_{i|t'} \Big)
    + \mD\Big(\rvf_{t}(\rvx_{t}) - \gamma \frac{\sqrt{\bar{\alpha}_t}}{\sqrt{1-\bar{\alpha}_t}} \sqrt{\rlambda_{i|t}} \rvv_{i|t'}\Big).
\end{equation}
This means effectively subtracting $\gamma \frac{\sqrt{\bar{\alpha}_t}}{\sqrt{1-\bar{\alpha}_t}} \sqrt{\rlambda_{i|t}} \rvv_{i|t'}$ from the noise prediction network output. 
Similar to \citet{haas2023discovering}, we empirically find that the difference between the two formulations only changes the amplification of the editing effect, and therefore opt to use a symmetric reverse process for simplicity.

Finally, we can write $\rvmu_{t}^\text{edit}$ explicitly by using both Eq.~(\ref{eq:Pdef}) and Eq.~(\ref{eq:Ddef}):
\begin{align}
    \rvmu_{t}^\text{edit}(\rvx_t) &= \sqrt{\bar{\alpha}_{t-1}} \Big(
    \frac{\rvx_t-\sqrt{1-\bar{\alpha}_t}\rvf_t(\rvx_t)}{\sqrt{\bar{\alpha}_t}}
    + \gamma \sqrt{\rlambda_{i|t}} \rvv_{i|t'}\Big) +
    \sqrt{1-\bar{\alpha}_{t-1} -\sigma_t^2}\Big( \rvf_t(\rvx_t) - \gamma \frac{\sqrt{\bar{\alpha}_t}}{\sqrt{1-\bar{\alpha}_t}} \sqrt{\rlambda_{i|t}} \rvv_{i|t'} \Big) \nonumber \\
    &= \sqrt{\bar{\alpha}_{t-1}} \Big(
    \frac{\rvx_t-\sqrt{1-\bar{\alpha}_t}\rvf_t(\rvx_t)}{\sqrt{\bar{\alpha}_t}}\Big)
    + \gamma \sqrt{\bar{\alpha}_{t-1}} \sqrt{\rlambda_{i|t}} \rvv_{i|t'} \nonumber \\
            &\qquad\qquad\qquad\qquad\qquad\qquad\qquad\quad + \sqrt{1-\bar{\alpha}_{t-1} -\sigma_t^2} \rvf_t(\rvx_t) - \gamma \sqrt{1-\bar{\alpha}_{t-1} -\sigma_t^2}\frac{\sqrt{\bar{\alpha}_t}}{\sqrt{1-\bar{\alpha}_t}} \sqrt{\rlambda_{i|t}} \rvv_{i|t'} \nonumber \\
    &= \sqrt{\bar{\alpha}_{t-1}} \mP\big(\rvf_t(\rvx_t)\big) + \mD\big(\rvf_t(\rvx_t)\big) 
      + \gamma \sqrt{\rlambda_{i|t}} \rvv_{i|t'} \Big(\sqrt{\bar{\alpha}_{t-1}} - \sqrt{1-\bar{\alpha}_{t-1} -\sigma_t^2}\frac{\sqrt{\bar{\alpha}_t}}{\sqrt{1-\bar{\alpha}_t}} \Big) \nonumber \\
      &= \rvmu_{t}(\rvx_t) + \gamma\sqrt{\rlambda_{i|t}}\rvv_{i|t'} \cdot \Big( \sqrt{\bar{\alpha}_{t-1}} - \frac{\sqrt{\bar{\alpha}_{t}}}{\sqrt{1-\bar{\alpha}_{t}}}\sqrt{1-\bar{\alpha}_{t-1} -\sigma_t^2}\Big).
\end{align}
Then, by setting 
\begin{equation}
    c_t=\sqrt{\bar{\alpha}_{t-1}} - \frac{\sqrt{\bar{\alpha}_{t}}
    }{\sqrt{1-\bar{\alpha}_{t}}}\sqrt{1-\bar{\alpha}_{t-1} -\sigma_t^2},
\end{equation}
we get Eq.~(\ref{eq:newDdpmMean}): 
\begin{equation}
    \rvx_{t-1}= \rvmu_t(\rvx_t) + \gamma c_t\rlambda_{i|t}^{1/2}\rvv_{i|t'} + \sigma_t \rvz_t,\quad t=T,\ldots,1. \nonumber
\end{equation}

PCs computed using subspace iterations for each timestep separately are calculated independently of one another. As such, PCs for adjacent timesteps might be highly correlated.
This is because in adjacent timesteps the noise level and uncertainty level are similar.
Specifically, PCs from adjacent timesteps might be highly negatively correlated, as the positive directions are independently chosen at each timesteps.
As explained in Sec.~\ref{sec:unsupervisedEditing}, one way of editing using Eq.~(\ref{eq:newDdpmMean}) involves setting $t'=t$, so that each denoising step is perturbed with its own PCs.
Therefore, it is possible that when perturbing the signal using PCs from neighboring timesteps they will cancel each other, thereby lessening the editing effect.
To that end, at the end of the PCs computation for each timestep we compare the current PCs, $\rvv_{i|t}$, to those calculated during the previous timestep, $\rvv_{i|t+1}$. 
When the PCs correlation is lower than some threshold $\rho<0$, we swap the direction of the current PCs, $\rvv_{i|t}$. 

In addition to publishing the code repository, we provide in Alg.~\ref{alg:unsup_pc_extraction} the complete algorithm for the unsupervised PC computation described here and in Sec.~\ref{sec:unsupervisedEditing} for reference.

\begin{algorithm}
    \caption{Unsupervised PCs Computation}
    \label{alg:unsup_pc_extraction}
    \begin{algorithmic}[1]
\STATE {\bfseries Inputs:} 
\STATE $\quad$ Timesteps to extract PCs for $\{T_\text{start}, \ldots, T_\text{end}\}$, 
\STATE $\quad$ Inverted noise vectors $\{\rvx_{T_\text{start}}, \rvz_{T_\text{start}},\ldots, \rvz_{T_\text{end}}\}$, 
\STATE $\quad$ Number of PCs $N$,
\STATE $\quad$ DDPM Denoiser $\rvf_t(\cdot)$, 
\STATE $\quad$ coefficients $\{\bar{\alpha}_{T_\text{start}},\ldots, \bar{\alpha}_{T_\text{end}}\}$, 
\STATE $\quad$ noise-levels $\{\sigma_{T_\text{start}}, \ldots, \sigma_{T_\text{end}}\}$, 
\STATE $\quad$ Threshold for correlation swap $\rho$, 
\STATE $\quad$ Approximation constant $C\ll 1$, 
\STATE $\quad$ Iterations amount $K$
\STATE
\STATE Initialize $\rvx_t \gets \rvx_{T_\text{start}}$
\FOR {$t \gets T_\text{start}$ to $T_\text{end}$}
\STATE $\rvx_{0|t} \gets \big(\rvx_t-\sqrt{1-\bar{\alpha}_t}\rvf_t(\rvx_t)\big)/\sqrt{\bar{\alpha}_t}$ $\qquad\qquad\qquad\;\;\;\,$\COMMENT {Run a normal reverse pass.}
\STATE $\rvmu_t \gets \sqrt{\bar{\alpha}_{t-1}} \rvx_{0|t} + \sqrt{1-\bar{\alpha}_{t-1} -\sigma_t^2}\rvf_t(\rvx_t)$
\STATE $\rvx_{t-1} \gets \rvmu_t + \sigma_t \rvz_t$
\STATE $\{\rvv_{i|t}^{(0)}\}_{i=1}^N \gets  \gN(0,\mI)$ $\qquad\qquad\qquad\qquad\qquad\qquad\:\:\:$\COMMENT {Extract PCs using $K$ subspace iterations over $\rvx_{0|t}$.}
    \FOR{$k \gets 1$ to $K$}                      
        \FOR{$i \gets 1$ to $N$}
            \STATE $\rvx_t^\text{shifted} \gets \rvx_t + C\sqrt{\bar{\alpha}_t} \rvv_{i|t}^{(k-1)}$ 
            \STATE $\rvx_{0|t}^\text{shifted} \gets \big(\rvx_t^\text{shifted}-\sqrt{1-\bar{\alpha}_t}\rvf_t(\rvx_t^\text{shifted})\big)/\sqrt{\bar{\alpha}_t}$
            \STATE $\rvv_{i|t}^{(k)} \gets \frac{1}{C}\big( \rvx_{0|t}^\text{shifted} - \rvx_{0|t} \big)$
        \ENDFOR
        \STATE $\mathbf{Q},\mathbf{R}\gets$ QR\_Decomposition($[\rvv_{1|t}^{(k)} \cdots \; \rvv_{N|t}^{(k)}]$) 
        \STATE $[\rvv_{1|t}^{(k)} \cdots \; \rvv_{N|t}^{(k)}] \gets \mathbf{Q}$
    \ENDFOR
    \STATE $\rvv_{i|t} \gets \rvv_{i|t}^{(K)}$ $\qquad\qquad\qquad\qquad\qquad\qquad\qquad\qquad\:$ \COMMENT{Save the computed PCs and EVs for timestep $t$.}
    \STATE $\lambda_{i|t} \gets \frac{1/\bar{\alpha}_t - 1}{C}\big\| \rvx_{0|t}^\text{shifted} - \rvx_{0|t}\big\|$
    \FOR{$i \gets 1$ to $N$}
        \IF{$\rvv_{i|t}\cdot \rvv_{i|t+1} < \rho$}
            \STATE $\rvv_{i|t} \gets - \rvv_{i|t}$ $\qquad\qquad\qquad\qquad\qquad\qquad\quad\;\;\;\,$ \COMMENT{Swap the PCs direction if it is highly negatively correlated \\ $\qquad\qquad\qquad\qquad\qquad\qquad\qquad\qquad\qquad\quad\quad\,$ with the previous PC.}
        \ENDIF
    \ENDFOR
\STATE $\rvx_t \gets \rvx_{t-1}$
\ENDFOR
    \end{algorithmic}
\end{algorithm}

\section{The Effect of Using PCs From Different Timesteps $\{t'\}$}\label{app:UnsupEntropy}

\begin{figure*}[t]
    \centering
    \includegraphics[width=\linewidth]{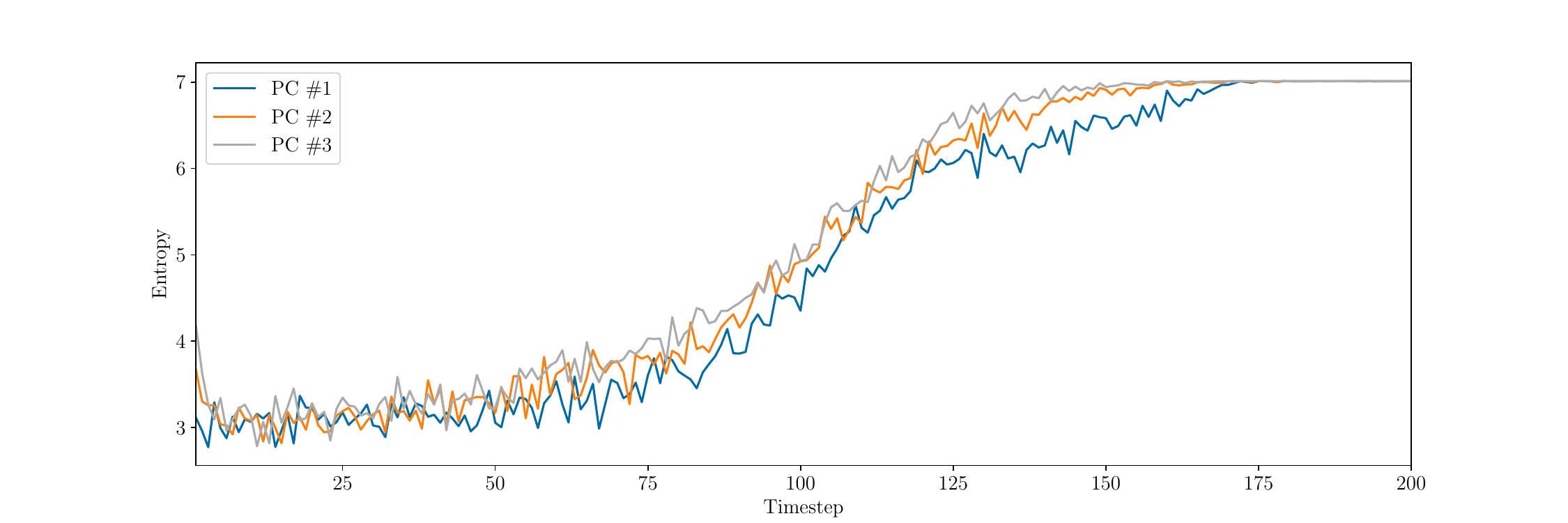}
    \caption{\textbf{Average entropy of the extracted PCs, across different timesteps.} Higher entropy is an indicator for global PCs, that change large segments of the signal.
    As the reverse process continues, the uncertainty in $\hat{\rvx}_{0|t}$ decreases, and the extracted PCs affect more localized areas, and measure lower in entropy.}
    \label{fig:unsupEntropy}
\end{figure*}

As mentioned in Sec.~\ref{sec:unsupExperiments}, modifications resulting from different choices of $\{t'\}$ vary in their extent of global impact.
This can be measured quantitatively by calculating the entropy of the PC when summing over the different channels and height of the tensors. Fig.~\ref{fig:unsupEntropy} displays the average entropy results for the first 3 PCs over the MusicDelta subset dataset used in the paper. Global changes that spread across large segments of the signal are characterized by a higher entropy, whereas lower entropy is an indicator for localized modifications. 
As can be seen, the entropy decays for smaller timesteps (later timesteps in the reverse diffusion process).

Empirically we see that the ability of the subspace iteration method to converge at large timesteps is hampered. This is also visible in Fig.~\ref{fig:unsupEntropy}, where the entropy is constant in the earliest timesteps.
The uncertainty at the start of the diffusion process is very large, which coincides with the existence of multiple PCs with similar strength, \ie there are no dominant directions. 
The subspace iteration method performs worse in such cases and is slower, and as such the extracted directions at  earlier timesteps are not very interesting. 
We also note that timesteps so early in the diffusion process are responsible for very global semantics, therefore editing them will result in a very large deviation from the original signal. Effectively, editing using those timesteps is equivalent to synthesizing a signal almost from scratch instead of editing it, which is not the desired task.
As the reverse process continues the uncertainty decreases, the signal's general structure is set, and the PCs display more fine-grained directions of editing.

\section{Comparison of Unsupervised Editing Directions With Random Directions}\label{app:randVector}

\begin{figure}[t]
    \centering
    \begin{subfigure}{0.5\linewidth}
        \centering
        \includegraphics[height=136pt]{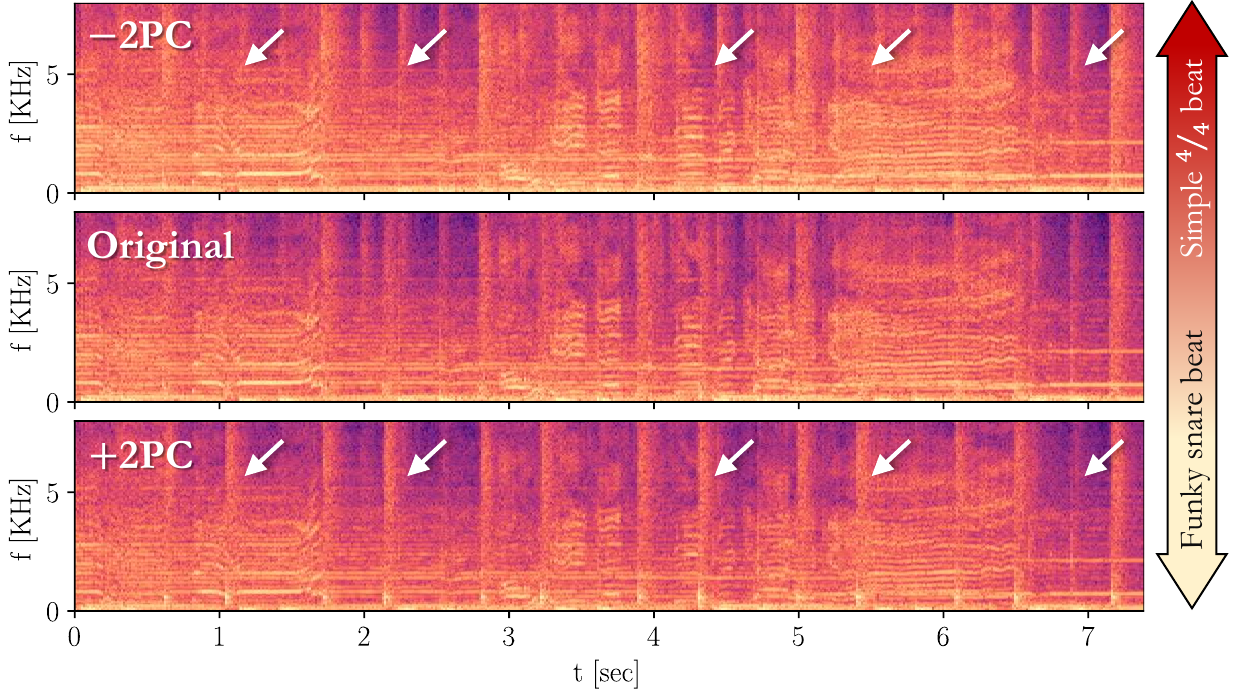}
        \caption{The first PC computed using our method, $\gamma=2$.}
        \label{fig:randv_ours_hendrix}
    \end{subfigure}%
    \begin{subfigure}{0.5\linewidth}
        \centering
        \includegraphics[height=136pt, trim={0 0 42pt 0}, clip]{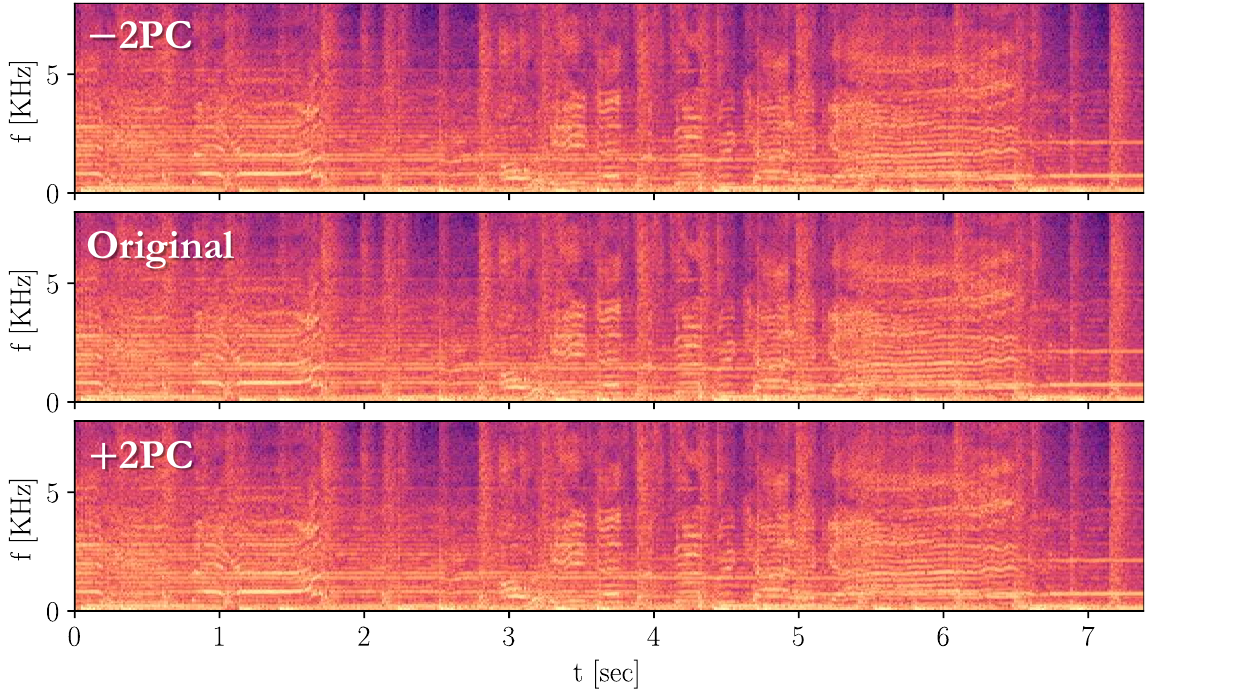}
        \caption{Random direction, $\gamma=2$.}
        \label{fig:randv_rand_hendrix2}
    \end{subfigure}\\
    \begin{subfigure}{0.5\linewidth}
        \centering
        \includegraphics[height=136pt]{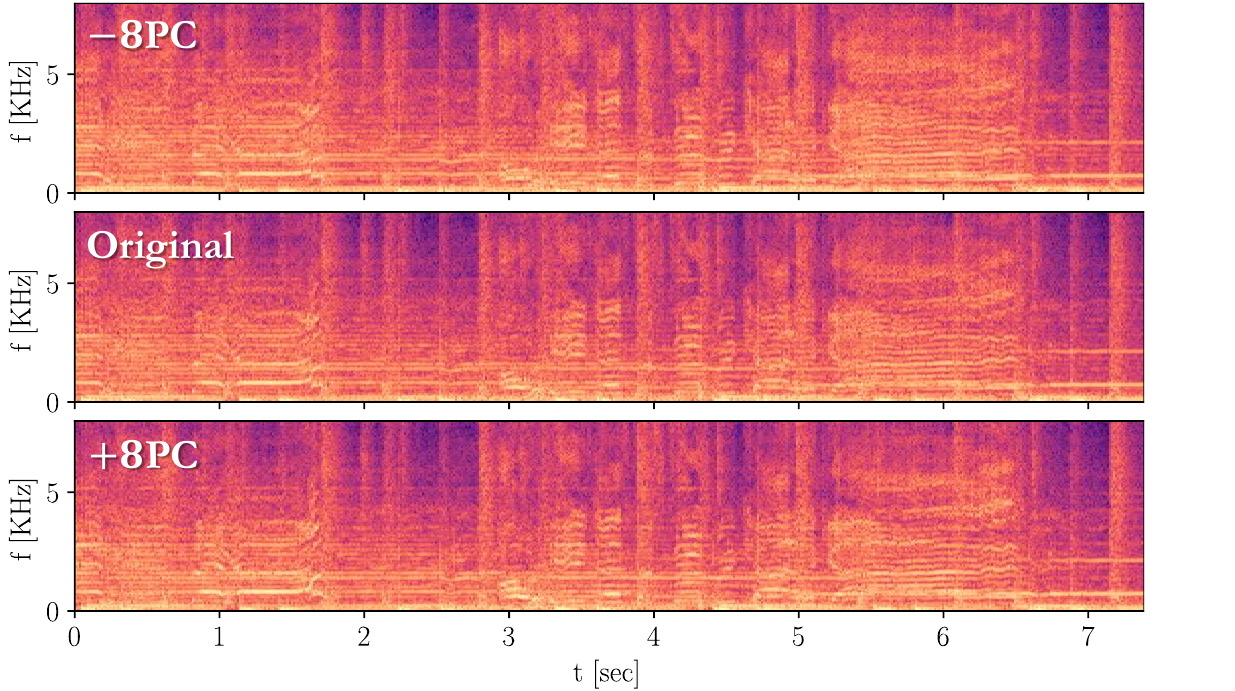}
        \caption{Random direction, $\gamma=8$.}
        \label{fig:randv_rand_hendrix8}
    \end{subfigure}%
    \begin{subfigure}{0.5\linewidth}
        \centering
        \includegraphics[height=136pt, trim={0 0 42pt 0}, clip]{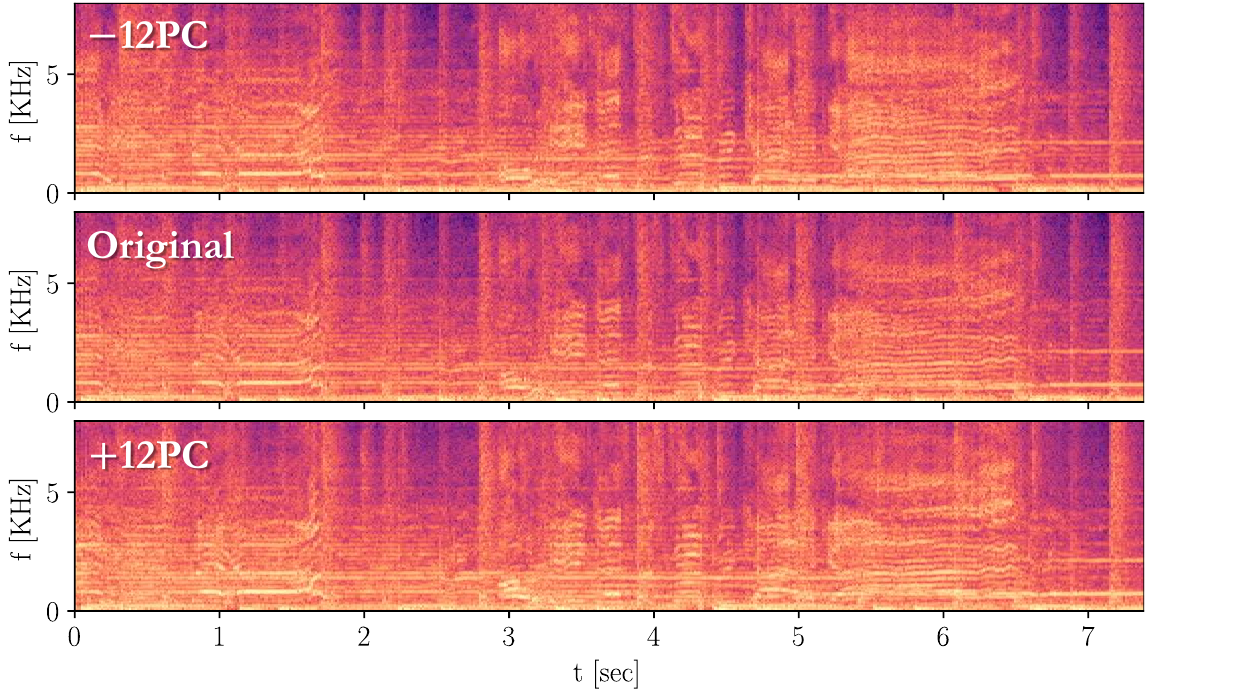}
        \caption{Random direction, $\gamma=12$.}
        \label{fig:randv_rand_hendrix12}
    \end{subfigure}
    \caption{\textbf{Our $1^{st}$ PC vs.~a random direction.}
    While our method extract PCs with a clear semantic meaning (\ref{fig:randv_ours_hendrix}), \eg changing the drums beat style from a simple 4/4 beat to a syncopated (off-beat) and snare-heavy beat, the effect of using random directions~(\ref{fig:randv_rand_hendrix2}),(\ref{fig:randv_rand_hendrix8}),(\ref{fig:randv_rand_hendrix12}) varies from producing unnoticeable changes with small $\gamma$ factor to degrading the signal when large $\gamma$ values are used. 
    This result can be listened to in \href{\urlofwebpagesupp\#randv}{Sec.~3 of our supplemental examples page}.
    We fix here $t'=80$ for our method, and apply our or the random directions for $T_\text{start}=200$, $T_\text{end}=1$.
    Both samples were generated using the music checkpoint of AudioLDM2, with the following source prompt randomly chosen from our prompts dataset: ``A high quality recording of a man singing with a rock band accompaniment.''}
    \label{fig:randv_hendrix}
\end{figure}

\begin{figure}[t]
    \centering
    \begin{subfigure}{0.5\linewidth}
        \centering
        \includegraphics[height=136pt]{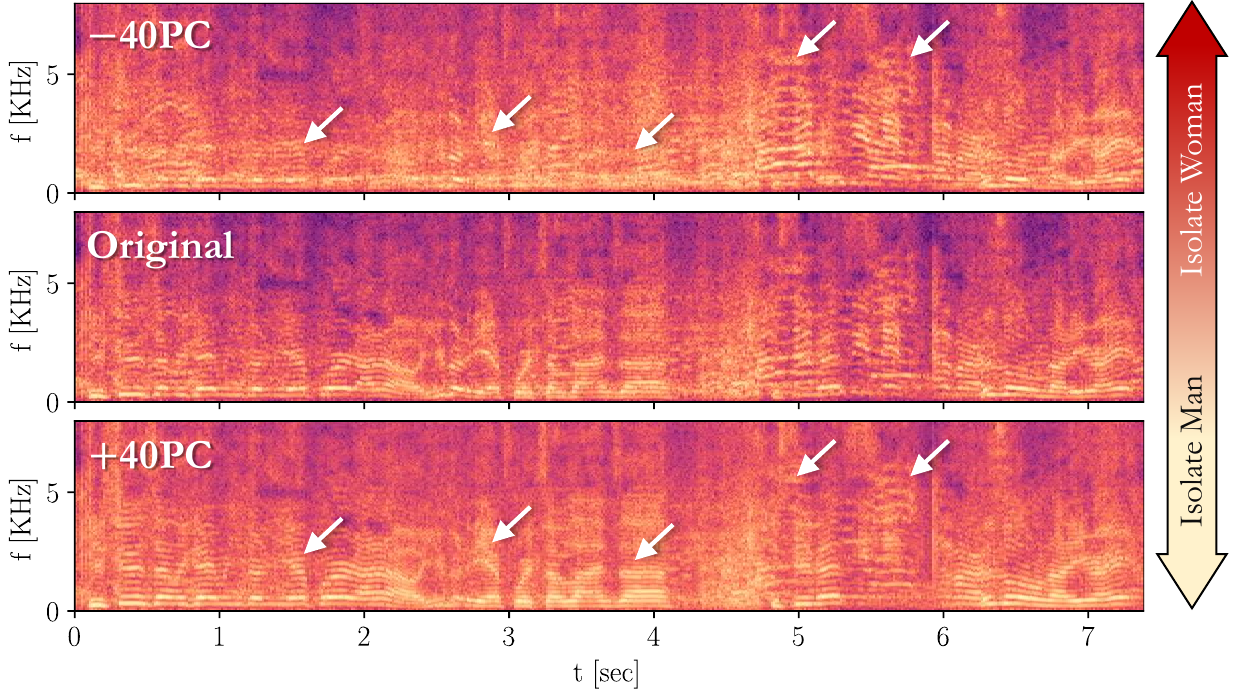}
        \caption{The first PC computed using our method, $\gamma=40$.}
        \label{fig:randv_ours_speech}
    \end{subfigure}%
    \begin{subfigure}{0.5\linewidth}
        \centering
        \includegraphics[height=136pt, trim={0 0 42pt 0}, clip]{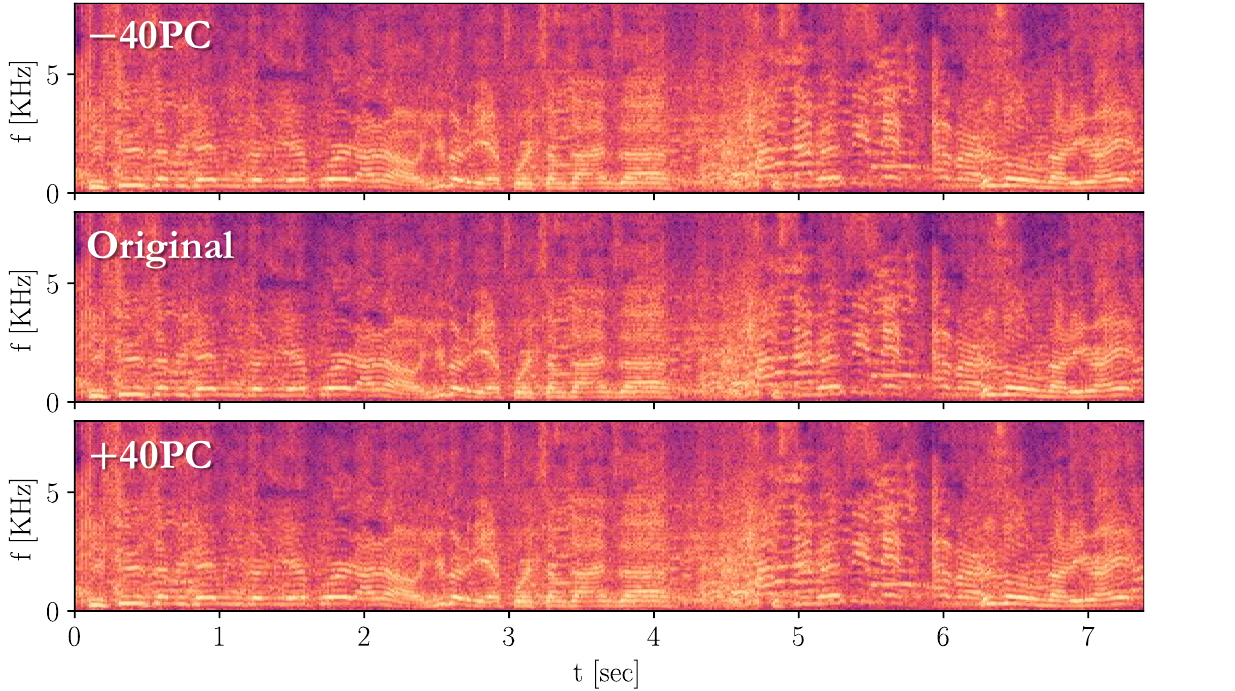}
        \caption{Random direction, $\gamma=40$.}
        \label{fig:randv_rand_speech40}
    \end{subfigure}\\
    \begin{subfigure}{0.5\linewidth}
        \centering
        \includegraphics[height=136pt]{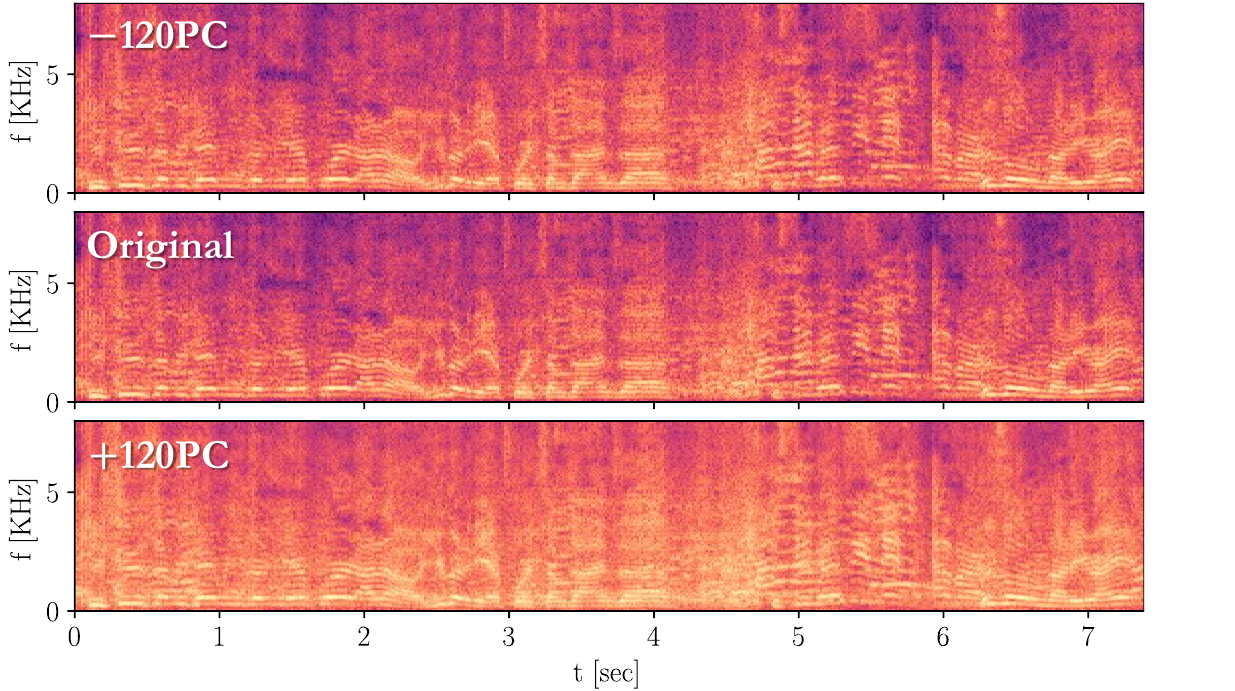}
        \caption{Random direction, $\gamma=120$.}
        \label{fig:randv_rand_speech120}
    \end{subfigure}%
    \begin{subfigure}{0.5\linewidth}
        \centering
        \includegraphics[height=136pt, trim={0 0 42pt 0}, clip]{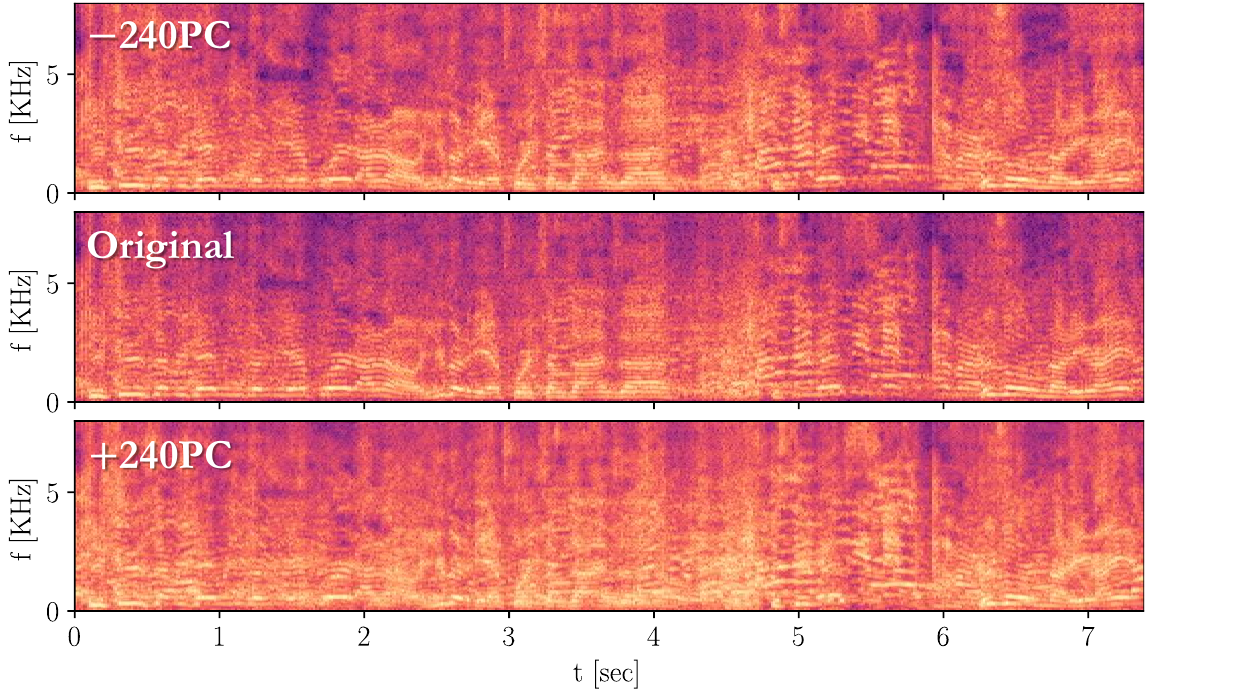}
        \caption{Random direction, $\gamma=240$.}
        \label{fig:randv_rand_speech240}
    \end{subfigure}
    \caption{\textbf{Our $1^{st}$ PC vs.~ a random direction.}
    While our method extract PCs with a clear semantic meaning (\ref{fig:randv_ours_speech}), \eg isolating a man or a woman speaking in a given signal, the effect of using random directions~(\ref{fig:randv_rand_speech40}),(\ref{fig:randv_rand_speech120}),(\ref{fig:randv_rand_speech240}) varies from producing unnoticeable changes with small $\gamma$ factor to degrading the signal when large $\gamma$ values are used. This result can be listened to in \href{\urlofwebpagesupp\#randv}{Sec.~3 of our supplemental examples page}.
    We fix $t'=t$, and apply our or the random directions for $T_\text{start}=115$, $T_\text{end}=95$.
    Both samples were generated using the large checkpoint of AudioLDM2, without a source prompt.}
    \label{fig:randv_speech}
\end{figure}

\begin{figure}[t]
    \centering
    \begin{subfigure}{0.5\linewidth}
        \centering
        \includegraphics[width=\linewidth]{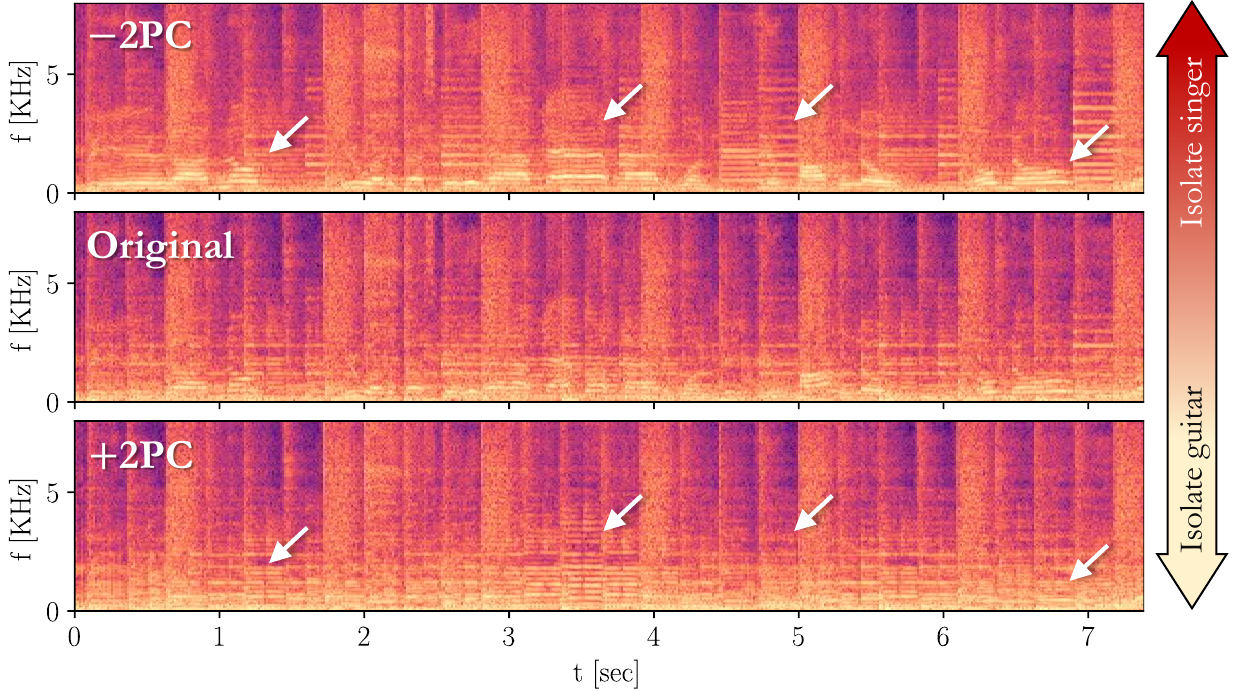}
        \caption{The first 3 PCs combined computed using our method, $\gamma=2$.}
        \label{fig:randv_ours_rock}
    \end{subfigure}%
    \begin{subfigure}{0.5\linewidth}
        \centering
        \includegraphics[height=136pt, trim={0 0 42pt 0}, clip]{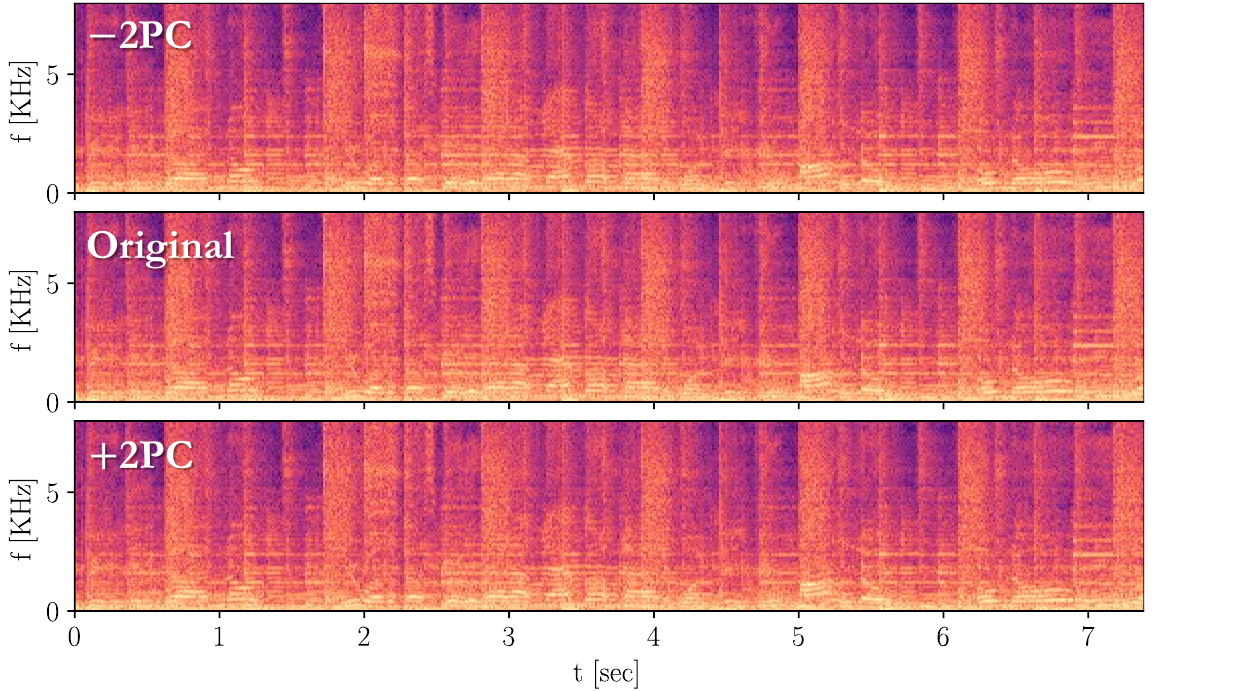}
        \caption{Random directions, $\gamma=2$.}
        \label{fig:randv_rand_rock2}
    \end{subfigure}\\
    \begin{subfigure}{0.5\linewidth}
        \centering
        \includegraphics[height=136pt]{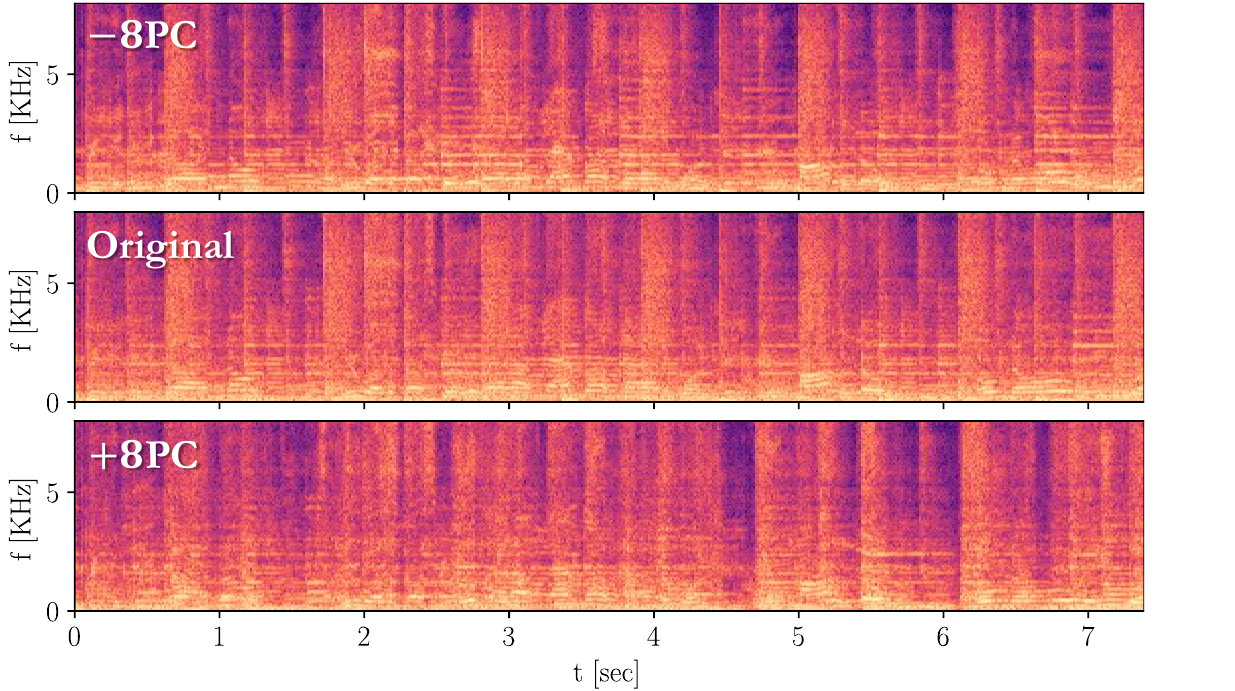}
        \caption{Random directions, $\gamma=8$.}
        \label{fig:randv_rand_rock8}
    \end{subfigure}%
    \begin{subfigure}{0.5\linewidth}
        \centering
        \includegraphics[height=136pt, trim={0 0 42pt 0}, clip]{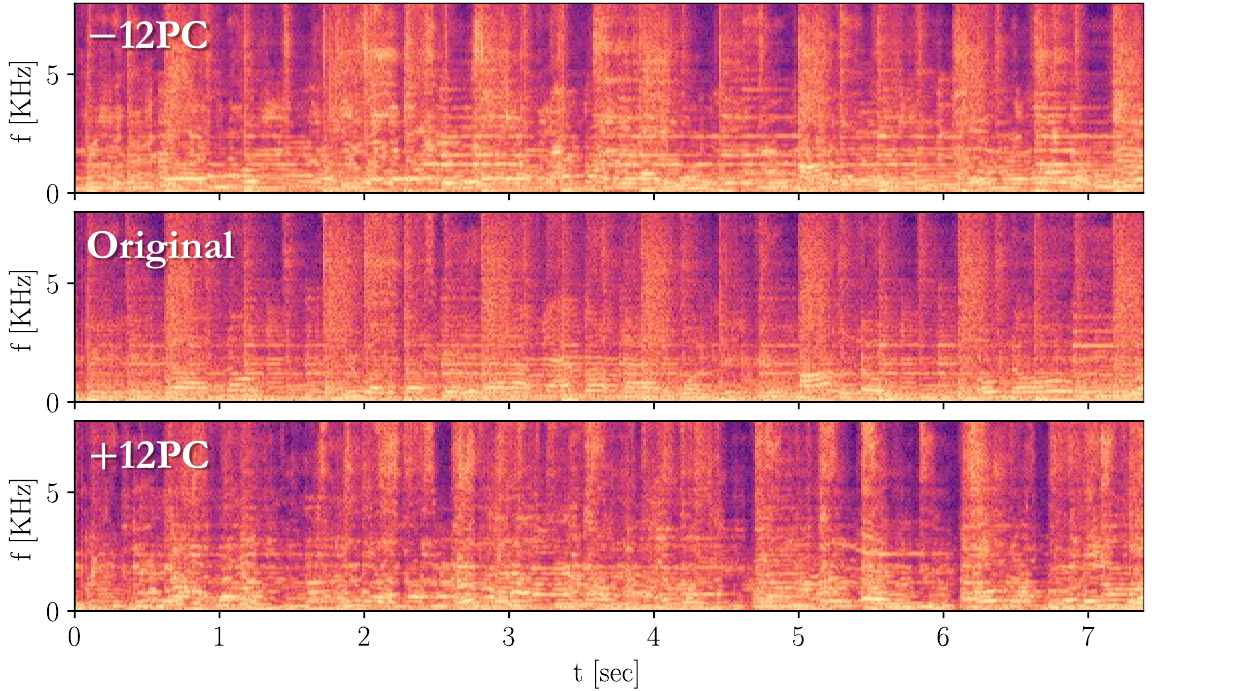}
        \caption{Random directions, $\gamma=12$.}
        \label{fig:randv_rand_rock12}
    \end{subfigure}
    \caption{\textbf{Our PCs vs.~random directions.}
    While our method extract PCs with a clear semantic meaning (\ref{fig:randv_ours_rock}), \eg isolating a singer or a guitar in a given signal, the effect of using random directions~(\ref{fig:randv_rand_rock2}),(\ref{fig:randv_rand_rock8}),(\ref{fig:randv_rand_rock12}) varies from producing unnoticeable changes with small $\gamma$ factor to degrading the signal when large $\gamma$ values are used. This result can be listened to in \href{\urlofwebpagesupp\#randv}{Sec.~3 of our supplemental examples page}.
    We fix $t'=65$ for our method, and apply our or the random directions for $T_\text{start}=200$, $T_\text{end}=1$.
    Both samples were generated using the music checkpoint of AudioLDM2, with the following source prompt randomly chosen from our prompts dataset: ``A recording of an old timey rock song from the sixties.''}
    \label{fig:randv_rock}
\end{figure}

To ensure our computed PCs are significant and do indeed carry semantic meaning, we compare them to random directions in Fig.~\ref{fig:randv_hendrix}, Fig.~\ref{fig:randv_speech} and Fig.~\ref{fig:randv_rock}.
Specifically, we compare for multiple signals the $1^{st}$ PC computed using our method, or a combination of the first 3 PCs, using both editing ways: (i) Applying a specific timestep $t'$ and adding it to a range of timesteps $[T_\text{start},\ldots,T_\text{end}]$, and (ii) setting $t'=t$ when adding the PCs to a range of timesteps.
When using a specific timestep, $t'$, across a range of timesteps, $[T_\text{start},\ldots,T_\text{end}]$, we sample a random direction from an isotropic Gaussian distribution. 
When settings $t'=t$, we randomly sample a direction for each timestep in the range $[T_\text{start},\ldots,T_\text{end}]$, sampled i.i.d from an isotropic Gaussian distribution.
In both cases, we normalize the randomly sampled directions to share the same norm as our directions, the unit-norm, and use the same computed eigenvalues $\lambda_{i|t}^{1/2}$ in Eq.~(\ref{eq:newDdpmMean}).

Our computed PCs display semantically meaningful editing directions, while using random directions over the strength $\gamma$ introduces almost imperceptible changes.
Using a large $\gamma$ factor introduces random changes that rapidly degrade the quality of the modification. All examples can be listened to in \href{\urlofwebpagesupp\#randv}{Sec.~3 of our supplemental examples page}.

\section{User Study}\label{app:userStudy}

\begin{figure}
    \centering
    \begin{subfigure}{\linewidth}
        \centering
        \includegraphics[width=0.9\linewidth]{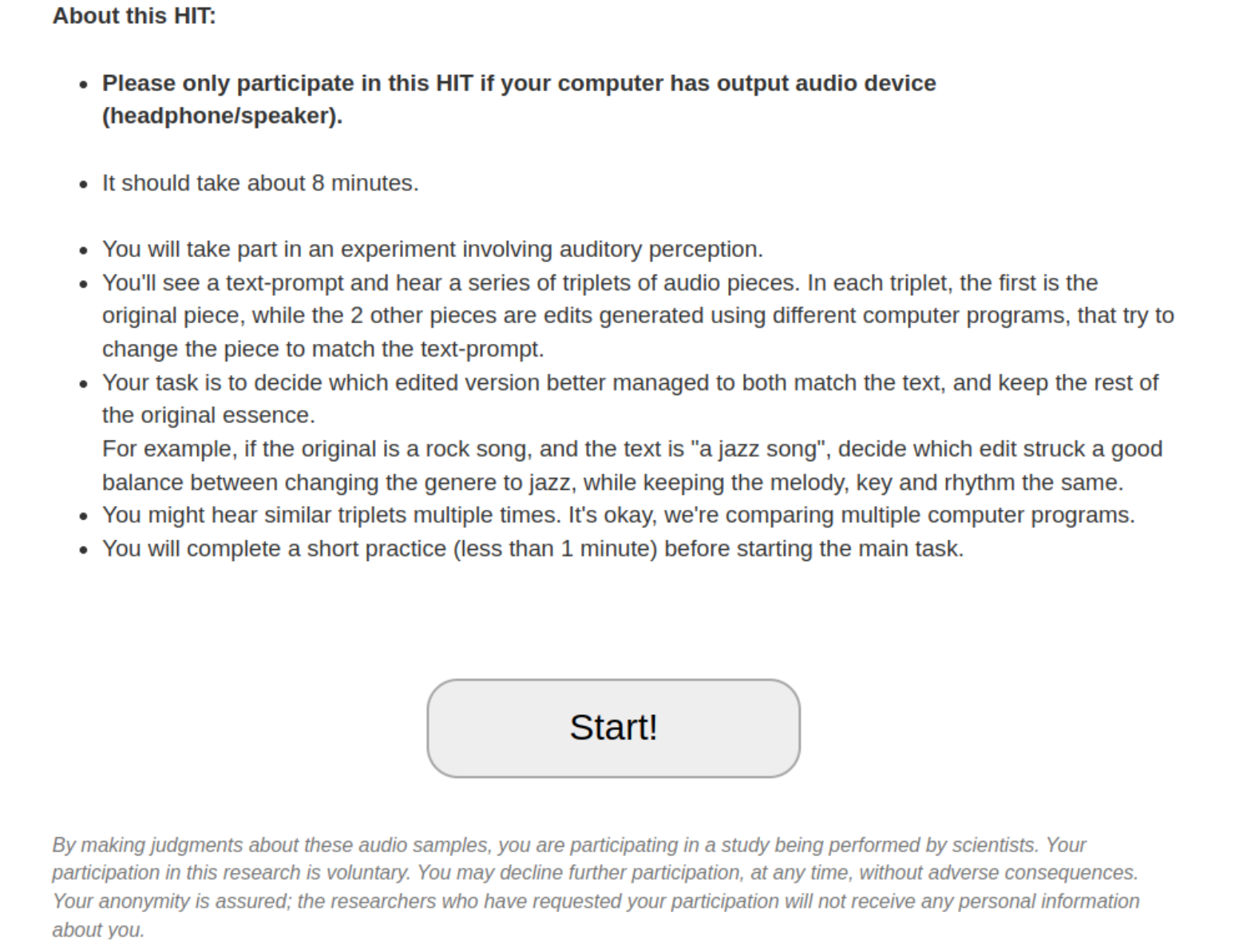}
        \caption{Instructions presented to a worker.}
        \label{fig:ustudy_inst}
    \end{subfigure}
    \begin{subfigure}{\linewidth}
        \centering
        \includegraphics[width=0.9\linewidth]{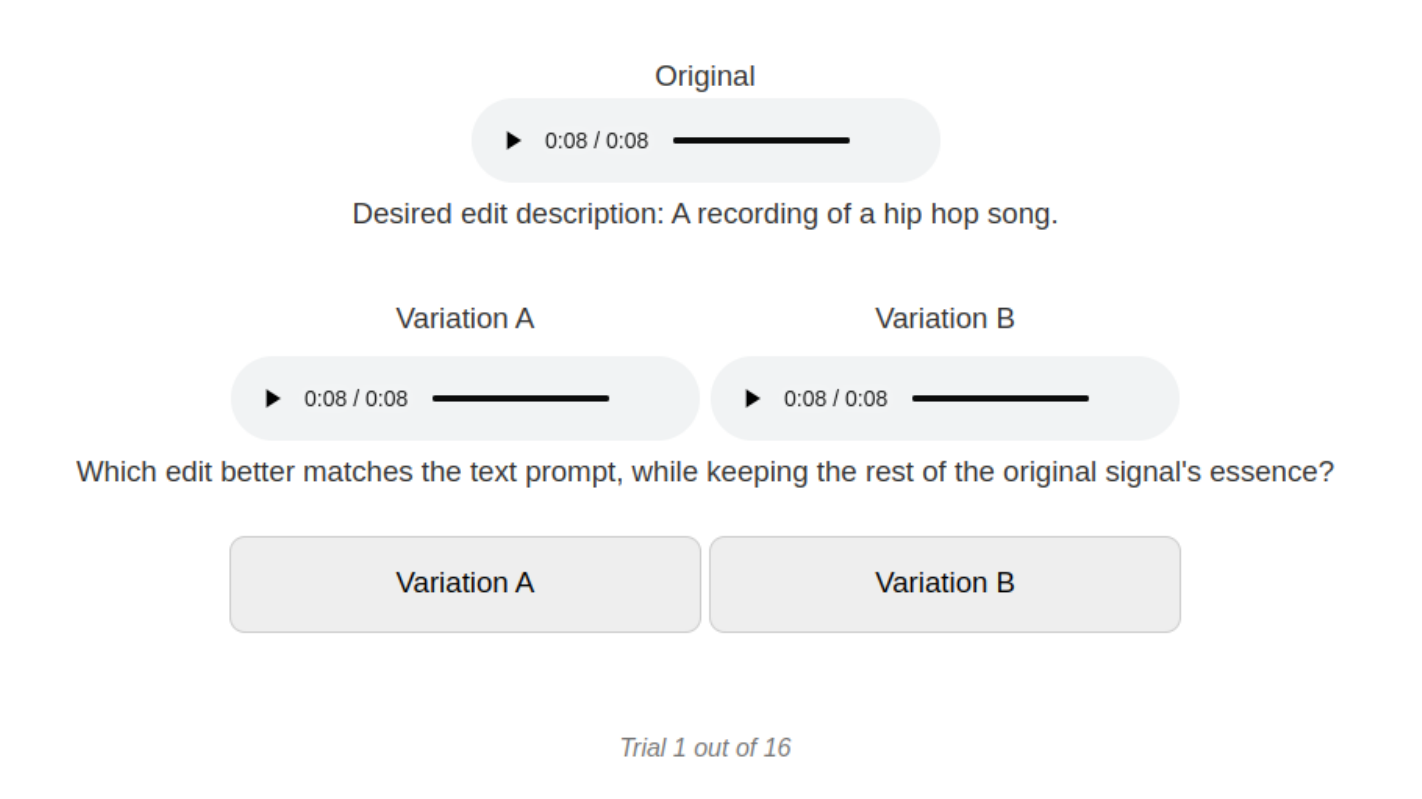}
        \caption{Example of a question screen.}
        \label{fig:ustudy_quest}
    \end{subfigure}
    \caption{\textbf{Screenshots from our user study.} After reading the instructions (upper), workers were shown 16 question screens (lower) to choose their preferred edit.}
    \label{fig:ustudy_screens}
\end{figure}

We include screenshots from our user study interface in Fig.~\ref{fig:ustudy_screens}. For fairness, all samples are normalized at the same loudness equal to $-19.11$ dBFS.

\section{Unsupervised Editing in Images}\label{app:images}

\begin{figure}
    \centering
    \includegraphics[width=\linewidth]{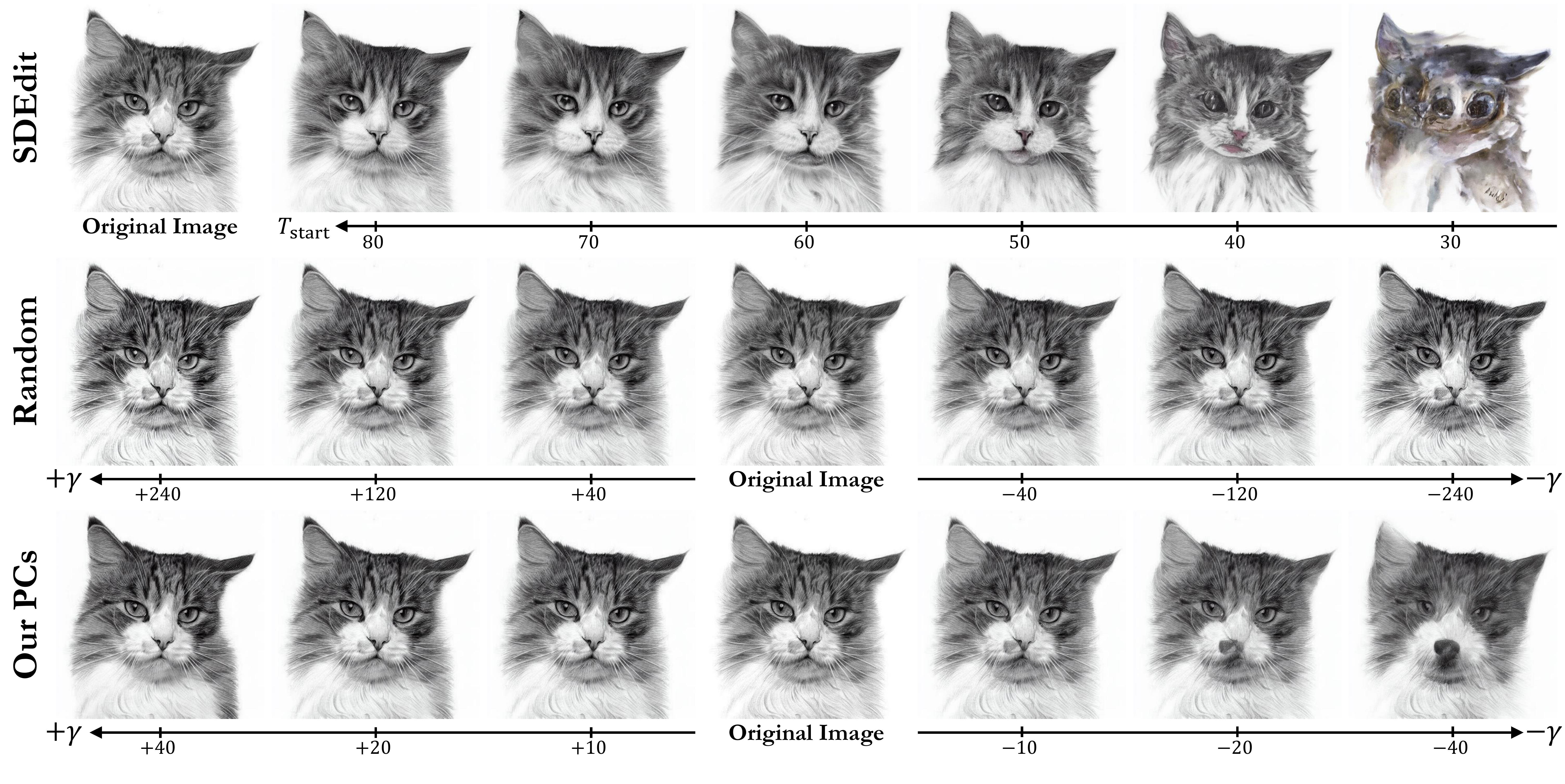}
    \caption{\textbf{Unsupervised zero-shot editing of images.} We demonstrate the applicability of our unsupervised approach for extract editing directions using Stable Diffusion~\citep{rombach2022high} on images taken from  ``modified ImageNet-R-TI2I''~\citep{tumanyan2023plug}, contrasted with SDEdit~\citep{meng2021sdedit} and our previously proposed random baselines (See App.~\ref{app:randVector}. Our method extracts editing directions that carry a semantic meaning, \eg a direction for changing the species of the cat or making it a more distinct cat-breed, while retaining the original essence of the image.}
    \label{fig:images_cat}
\end{figure}

\begin{figure}
    \centering
    \includegraphics[width=\linewidth]{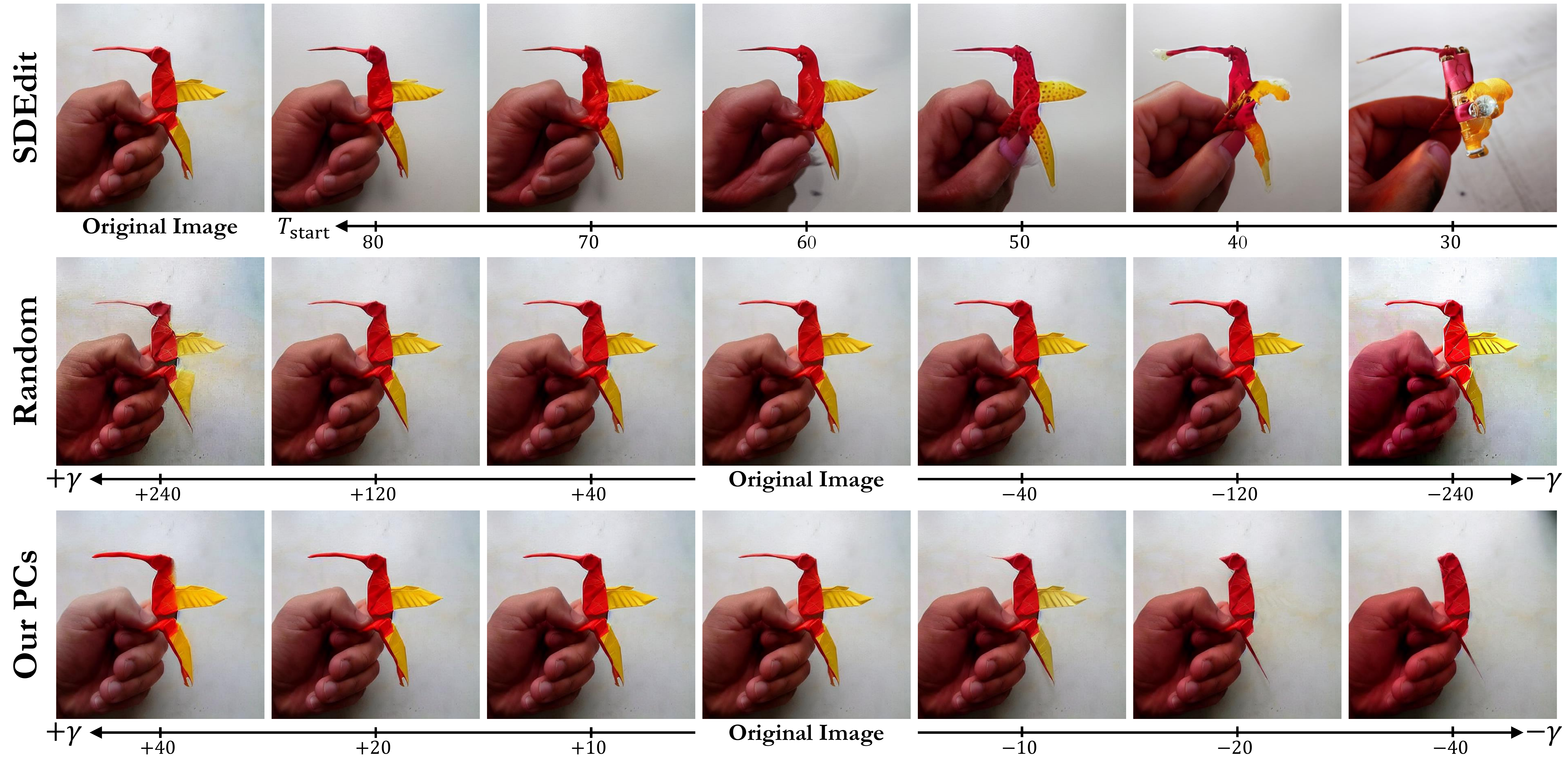}
    \caption{\textbf{Unsupervised zero-shot editing of images.} We demonstrate the applicability of our unsupervised approach for extract editing directions using Stable Diffusion~\citep{rombach2022high} on images taken from  ``modified ImageNet-R-TI2I''~\citep{tumanyan2023plug}, contrasted with SDEdit~\citep{meng2021sdedit} and our previously proposed random baselines (See App.~\ref{app:randVector}. Our method extracts editing directions that carry a semantic meaning, \eg thickening the beak of a bird or shortening it until it is no longer a bird and the wings dissapear, while retaining the essence and structure of the image, \eg not degrading the hand.}
    \label{fig:images_bird}
\end{figure}

\begin{figure}
    \centering
    \includegraphics[width=\linewidth]{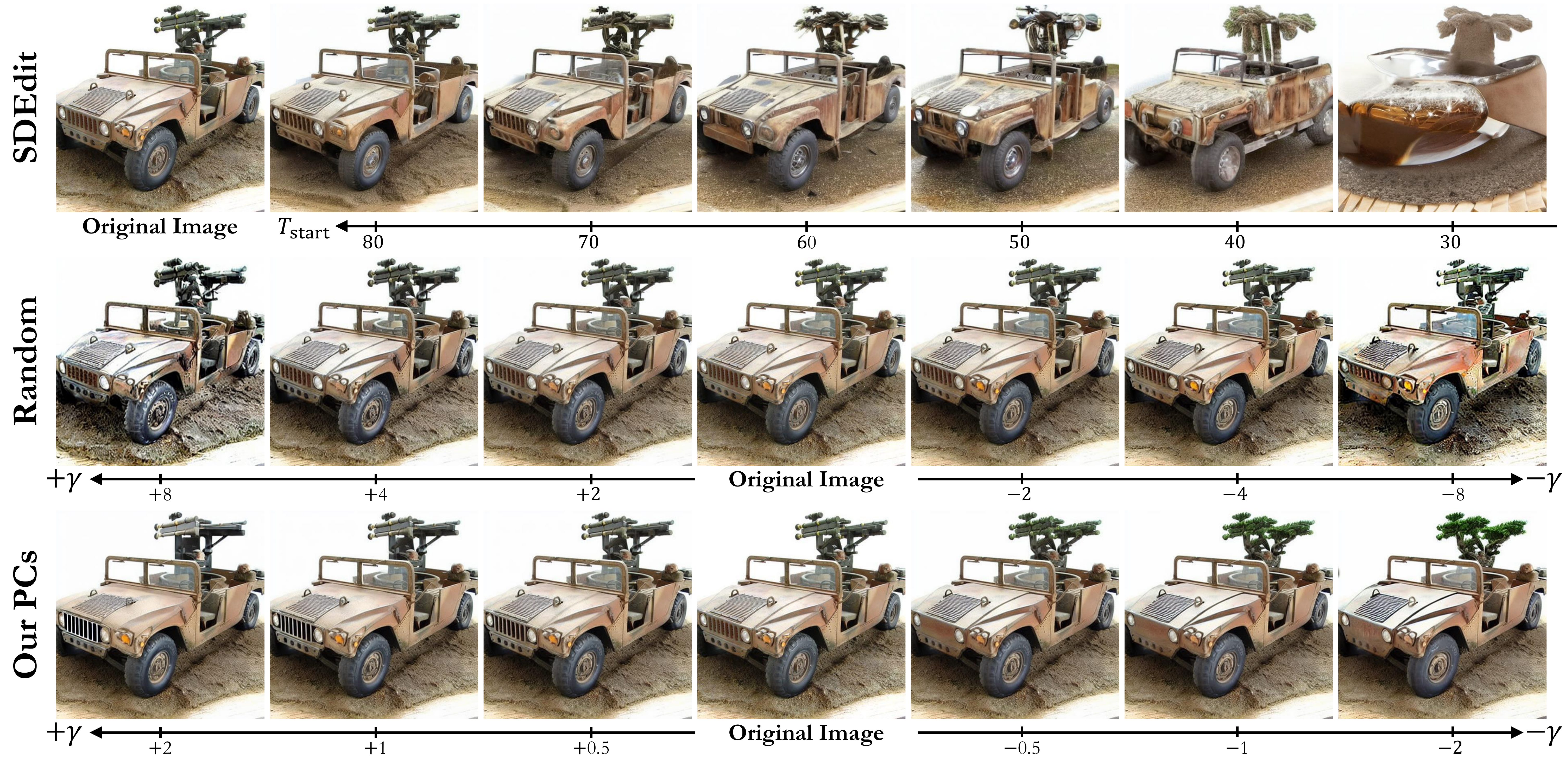}
    \caption{\textbf{Unsupervised zero-shot editing of images.} We demonstrate the applicability of our unsupervised approach for extract editing directions using Stable Diffusion~\citep{rombach2022high} on images taken from  ``modified ImageNet-R-TI2I''~\citep{tumanyan2023plug}, contrasted with SDEdit~\citep{meng2021sdedit} and our previously proposed random baselines (See App.~\ref{app:randVector}. Our method extracts editing directions that carry a semantic meaning, \eg turning a gun to a bonsai tree, while retaining the essence and structure of the image, \eg keeping the jeep toy car relatively the same.}
    \label{fig:images_jeep}
\end{figure}

We demonstrated our novel unsupervised editing approach on audio signals, since when applied on music it exposes a range of musically interesting modifications, some of which are virtually impossible to describe by text precisely.
However, this approach is not limited to the audio domain. 
As a preliminary demonstration of its generalization ability, we leverage Stable Diffusion~\citep{rombach2022high} to demonstrate in Figs.~\ref{fig:images_cat}, \ref{fig:images_bird} and Fig.~\ref{fig:images_jeep} semantic editing directions extracted for images from ``modified ImageNet-R-TI2I''~\citep{tumanyan2023plug}, and contrast the results with both the well known SDEdit~\citep{meng2021sdedit} and our previously proposed random baseline (See App.~\ref{app:randVector}). In all of the examples, no source prompts were used.
Our editing directions encode meaningful direction that change semantic elements while keeping the rest of the images, both its essence (\eg a sketch of a cat) and its structure intact. 

\end{document}